\documentclass[traditabstract,a4paper]{aa}
\usepackage{graphicx}
\usepackage{url}
  
\usepackage{times}
\newcommand{\MSun}{\mbox{M$_\odot$}}
\newcommand{\LSun}{\mbox{L$_\odot$}}
\newcommand\patspeed{\ensuremath{\text{km\,s}^{-1}\,\text{kpc}^{-1}}}

\def\apgt{\ {\raise-.5ex\hbox{$\buildrel>\over\sim$}}\ }
\def\aplt{\ {\raise-.5ex\hbox{$\buildrel<\over\sim$}}\ }
\def\lteq{\ {\raise-.5ex\hbox{$\buildrel<\over-$}}\ }

\newcommand{\Al}{$^{26}$Al\,}
\newcommand{\Fe}{$^{60}$Fe\,}

\usepackage{color}

\begin{document}

\title{The Formation of Solar System Analogs in Young Star Clusters}

\author{S.\, Portegies Zwart\inst{1}}

\offprints{S. Portegies Zwart}
\mail{spz@strw.leidenuniv.nl}
\institute{
 $^1$Leiden Observatory, Leiden University, PO Box 9513, 2300
   RA, Leiden, The Netherlands
}

\date{Received / Accepted }
\titlerunning{Supernova Near the Solar System}
\authorrunning{Portegies Zwart et al.}

\abstract{
  
The Solar system was once rich in the short-lived radionuclide (SLR)
\Al\, but deprived in \Fe. Several models have been proposed to
explain these anomalous abundances in SLRs, but none has been set
within a self-consistent framework of the evolution of the Solar
system and its birth environment.  The anomalous abundance in \Al\,
may have originated from the accreted material in the wind of a
massive $\apgt 20$\,\MSun\, Wolf-Rayet star, but the star could also
have been a member of the parental star-cluster instead of an
interloper or an older generation that enriched the proto-solar
nebula.  The protoplanetary disk at that time was already truncated
around the Kuiper-cliff (at $45$\,au) by encounters with another
cluster members before it was enriched by the wind of the nearby
Wolf-Rayet star.  The supernova explosion of a nearby star, possibly
but not necessarily the exploding Wolf-Rayet star, heated the disk to
$\apgt 1500$\,K, melting small dust grains and causing the
encapsulation and preservation of \Al\, into vitreous droplets.  This
supernova, and possibly several others, caused a further abrasion of
the disk and led to its observed tilt of $5.6\pm1.2^\circ$ with
respect to the Sun's equatorial plane. The abundance of \Fe\,
originates from a supernova shell, but its preservation results from a
subsequent supernova. At least two supernovae are needed (one to
deliver \Fe\, and one to preserve it in the disk) to explain the
observed characteristics of the Solar system.  The most probable birth
cluster then has $N = 2500\pm300$\, stars and a radius of $r_{\rm vir}
= 0.75\pm0.25$\,pc.  We conclude that Solar systems equivalent systems
form in the Milky Way Galaxy at a rate of about 30 per Myr, in which
case approximately 36,000 Solar system analogues roam the Milky Way.

}

\maketitle

\section{Introduction}

There are several observables which make the Solar system at odds with
other planetary systems \citep{2004MNRAS.354..763B}. Apart from the
Solar system's planetary topology, these include the curiously small
disk of only $\sim 45$\,au, the morphology at the outer edge, and the
tilt of the ecliptic with respect to the equatorial plane of the
Sun. The high abundance of \Al\, with respect to the Galactic
background also seems odd.  Each of these observables may be the
result of the early evolution of the Solar system. It is controversial
to think that the Solar system is really different than other
planetary systems \citep{1632Dialogo...G,1922ApJ....55..302K}, and
naively one would expect that its characteristics are a natural
consequence of the environment in which it.
Let's start with the curious abundance of \Al/$^{27}$Al$ = 4.5$---$5.2
\times 10^{-5}$ as observed today in Calcium-Aluminum Inclusions
(CAIs) and vitreous chondrules
\citep{1995Metic..30..365M,2008LPI....39.1999J}.  These solids formed
at temperatures of $\apgt 1500$\,K \citep{1990Metic..25..309H}, but
they have a much lower abundance in \Fe\, of only \Fe/$^{56}$Fe$ = 3.8
\pm 6.9 \times 10^{-8}$ \citep{2018ApJ...857L..15T}. The ratio of
\Fe/\Al\, excludes an origin from a nearby core-collapse supernova
explosion because this would result in a very low
\citep{2010ApJ...711..597O} but comparable abundances in \Al\, as well
as in \Fe\, \citep{2006NuPhA.777..424N}.  An even earlier enrichment
of the pre-solar nebula by the wind of a 1.6 to 6\,M$_\odot$\,
asymptotic giant-branch star
\citep{2000A&A...361..959M,2006NuPhA.777....5W} is hard to reconcile
with the timescales of star formation and disk evolution
\citep{2009ApJ...701..260I}, and an early pre-solar enrichment through
a $\apgt 20$\,M$_\odot$\, Wolf-Rayet star
\citep{1988ApJ...332..305D,2009ApJ...696.1854G,2010ApJ...714L..26T,gounelle12,2017ApJ...851..147D}
and its subsequent supernova would lead to an anomalously high
abundance in \Fe.  These scenarios have difficulty explaining the
observed SLRs and neither of these explains the outer edge of the
solar system's planetesimal disk, its tilt with respect to the Sun's
equatorial plane or the high temperatures needed for producing
vitreous droplets in chondrules. The alternatives to the latter, such
as electric discharge \citep{HORANYI1995174} and asteroidal collisions
\citep{ARAKAWA2016102}, are controversial \citep{SANDERS_2012}.

Instead of enriching the molecular cloud before the Sun formed, maybe
the parental cluster hosted a Wolf-Rayet star. This star may have
enriched the Sun's proto-planetary disk directly by accretion from its
copious stellar wind before it exploded in a supernova.  The hosting
stellar cluster has to be sufficiently massive ($\apgt 500$\,\MSun) to
assure a Wolf-Rayet star to be present and sufficiently dense to have
its wind enrich the proto-planetary disks through accretion.
Wolf-Rayet stars require a few Myr before they develop a massive
\Al-rich wind, and in a dense environment the majority of the
proto-planetary disks will by that time already have been truncated
severely by stellar encounters
\citep{2014MNRAS.444.2808P,2016MNRAS.457..313P,2016ApJ...828...48V} or
they may have evolved to a transient disk \citep{2015A&A...576A..52R}.
Such truncation would be a natural consequence of a dense
birth-cluster \citep{2009ApJ...696L..13P}, and it is reconcilable with
the short half-life for stars with disks in a clustered environment
\citep{2018MNRAS.477.5191R}.  The expectation is that disks which are
affected in the early cluster evolution eventually evolve into
planetary systems comparable to that of the Sun
\citep{2017MNRAS.471.2753R}.

We take these effects, the truncation of the disk due to close stellar
encounters and the accretion of \Al-enriched material from a
Wolf-Rayet wind, and the effect of nearby supernovae into account in
simulations of the Sun's birth cluster. Disks in our calculations tend
to be truncated considerably even before they can be enriched through
accreting material from the wind of a Wolf-Rayet star.  From the
moment the SLRs are released from the surface of the Wolf-Rayet star,
they start to decay.  The accretion onto a circum-stellar disk will
not prevent the further decay of SLRs.  In the Solar system, the
left-over by-products of \Al\, are found in vitreous droplets and
CAIs. These form at temperatures $\apgt 1500$\,K
\citep{2005mcp..book..407D}.  Such high temperatures are sufficient to
melt the dust particles and encapsulates earlier accreted \Al\, into
vitreous droplets.  Such high temperatures could be the result of a
nearby supernova that has irradiated the disk.  A supernova would
therefore provide a natural means to embed the SLRs in vitreous
droplets.  The Wolf-Rayet star that initially delivered the \Al\,
could be responsible for preserving the SLRs when it explodes at an
age of $3$ to $9$\,Myr \citep[for a $60$ to $20\,M_\odot$\, star of
  solar composition][]{2004NewAR..48....7V}, but it could also have
been another subsequent supernova.

In order to heat a circum-stellar disk to $\apgt 1500$\,K, the
supernova has to be in close proximity.  Such a nearby supernova has
considerable consequences for the further evolution of the
circumstellar disk. Apart from being heated, the protoplanetary disk
is also harassed by the nuclear blast-wave of the supernova
shell. This may lead to the truncation of the disk through
ram-pressure stripping \citep{2018arXiv180204360P}, and induces a tilt
to the disk due to the hydrodynamical-equivalent of the Stark effect
\citep{2017A&A...604A..88W}.  Both processes may be responsible for
shaping the outer edge of the Solar system, truncating it at about
45\,au and tilting the disk with respect to the Sun's equatorial
plane of $i_{\rm disk} = 5.6\pm1.2^\circ$.

The supernova blast-wave is insufficiently dense to copiously enrich
the surviving proto-planetary disk with SLRs produced in the exploding
star \citep{2007ApJ...662.1268O,2010ApJ...711..597O}, which is
consistent with the low abundance in observed \Fe. In addition, the
accreted \Fe\, decays and the information of its induced abundance can
only be preserved when, just like the accreted \Al\, also the \Fe\, is
captures in vitreous droplets. This requires a second heating of the
disk to $\apgt 1500$\,K.

The chain of events to explain the size of the disk, its tilt and the
abundances of \Al\, and \Fe, requires a windy Wolf-Rayet star and two
supernovae in short succession.  This seems extraordinary but appears
to be a natural consequence of being born in a cluster.  Here we
discuss this chain of events and how it is to be reconcilable with the
Solar system's birth environment.  We quantify these processes by
means of simulations in which we take the effects which naturally
follow from the sun being born in a star cluster into account. These
effects include the truncation of the circum-stellar disk due to close
stellar encounters, the accretion of \Al\, enriched material from the
copious wind of a nearby Wolf-Rayet star and the effects of nearby
supernovae. It turns out that these processes lead naturally to a
reasonably high formation rate of systems with characteristics similar
to the Solar system. We further constrain the fundamental parameters
of the clusters in which the Solar system was born and argue that the
environment from which the Solar system is most likely to have emerged
was a reasonably rich star cluster of moderate density.

\section{The computational approach}

\subsection{The Sun's clustered birth envionment}

We expand the analysis of \cite{2018arXiv180204360P} by simulating the
evolution of young clusters of 50 to $10^4$ stars.  The integration of
the equations of motion is performed in the potential of a two-armed
spiral galaxy with a bar
\citep{2016MNRAS.457.1062M,2017MNRAS.464.2290M}, and we keep track of
the disk truncation and mass loss due to close stellar encounters (see
\S\,\ref{Sect:Nbody}), the Bondi-Hoyle accretion of \Al\, from
Wolf-Rayet winds (see \S\,\ref{Sect:WRWinds}), and the effect of
supernova explosions on the protoplanetary disks (see
\S\,\ref{Sect:supernova}).  Each time a supernova heats a
protoplanetary disk to a temperature of $>1500$\,K, we preserve its
present composition by stopping the decay of previously accreted
SLRs. Multiple supernovae may then lead to multiple epochs of
preservation with a different relative composition. When
insufficiently heated, nuclear decay continues to reduce the
concentration of SLRs in the disk.  After the last supernova explosion
occurred, at an age of $\aplt 50$\,Myr, any disk that was not
preserved hardly shows traces of SLRs; only sufficiently heated disks
show high abundances, considerable truncation and a finite tilt angle
with respect to the initial orientation of the disk.

\subsection{The numerical procedure}

The simulations are performed using the Astrophysical Multipurpose
Software Environment \citep[AMUSE for short, see
  \S\,\ref{Sect:AMUSE}][]{PortegiesZwart2013456,2013AA...557A..84P,2011ascl.soft07007P}.
The calculations are separated into three distinct parts, in which we
simulate
\begin{itemize}
  \item the effect of the supernova irradiation on a nearby
    protoplanetary disk (\S\,\ref{Sect:Irradiation}),
  \item the effect of the supernova blast wave (\S\,\ref{Sect:blastwave}),
  \item and the consequences of encounters, accretion from stellar
    winds and supernovae on protoplanetary disks in young star
    clusters (this \S).
\end{itemize}
The parametrized effects of the supernova irradiation and blast-wave
impact are integrated together with the equations of motion for the
stars in the cluster.  By the time a star loses mass in a wind or
explodes in a supernova the effect on the other stars is taken into
account. In these calculations, the stellar mass-loss parameters and
supernovae are provided by the staller evolution code, whereas the
masses, positions and relative velocities or the stars with respect to
the mass-losing star are provided by the $N$-body code.  We perform a
grid of calculations in cluster density, structure, mass and
virialization to find the parameters for which it is most probable to
form a planetary system with characteristics (stellar mass and disk
size, mass, and inclination) similar to the early Solar system.  In
the Appendix, we briefly discuss the various ingredients in the
simulations. All calculations are performed to an age of 50\,Myr,
after which we analyze the results in searching for Solar system
analogues.

\subsection{Initial conditions}

Each simulation starts by selecting the masses, positions and
velocities of $N$ stars.  The stars in the clusters are distributed
according to a \cite{1911MNRAS..71..460P} sphere density distribution
or using a fractal with the box-counting dimension $F=1.6$
\citep{2004A&A...413..929G}.  The stars are selected randomly from a
broken power-law initial mass-function \citep{2002Sci...295...82K}
between the hydrogen-burning limit and an upper limit according to
\citep[][, in the discussion \S\, we relax this
  assumption]{2003ApJ...598.1076K}.

The size of the cluster is characterized with a virial radius of
$r_{\rm vir} = 0.1$\,pc, 0.3, 0.5, 0.7, 1.0, 1.5, 2.0, 2.5, 3.0 and
4\,pc.  The number of stars in each cluster is $N=900$ (the smallest
number which is still expected to host a Wolf-Rayet star, see
\S\,\ref{Sect:IMF}), 1000, 1500, 2000, 2500, 3000, 3500, 3500, 6000
and 10000.  In \S\,\ref{Sect:Discussion} we discuss the consequences
of using a mass function with a fixed uppper mass-limit of
120\,\MSun\, in which case also lower mass clusters, down to about
50\,\MSun\, are able to host a Wolf-Rayet star.

For each combination of $r_{\rm vir}$ and $N$ we perform
10 simulations up to an age of 50\,Myr, which is well beyond the
moment of the last supernova in the cluster, and which assures that
non-preserved SLRs have decayed to an unmeasurably small abundance. In
addition to the cluster mass and size we also varied the initial
virial rato from $q=0.4$, 0.5 (virial equilibrium), 0.7 and $q=1.3$.

At birth, each star receives a 400\,au radius disk with a mass of 10\%
of the stellar mass in a random orientation around the parent star.
This disk size is close to the upper limit of circum-stellar disks
observed using Atacama Large Millimetre Array in the nearby open
star-clusters in Lupus \citep{2017A&A...606A..88T}, Upper-Scorpius
\citep{2017ApJ...851...85B}, Ophiuchus \citep{2017ApJ...845...44T},
Orion \citep{2018ApJ...860...77E} and Taurus
\citep{2017ApJ...845...44T}.  During the integration of the equations
of motion, we follow the mass and size evolution of the disks, in
particular, the truncation due to encounters from passing other stars
(see Appendix \S\,\ref{Sect:DiskRadius}), their enrichment by the
winds of Wolf-Rayet stars (\S\,\ref{Sect:WRWinds}), and the influence
of nearby supernovae (\S\,\ref{Sect:supernova}).

\subsubsection{Selecting masses from the initial mass-function}\label{Sect:IMF}

Masses are selected from a broken power-law
mass function \citep{2002Sci...295...82K} between 0.08\,\MSun\, and an
upper limit depending on the total cluster mass according to
\citep{2006MNRAS.365.1333W}.  We approximate this upper mass limit with
\begin{eqnarray}
  \log_{10}\left( {m_{\rm max} \over \MSun} \right) &=& -0.76 
  + 1.06 \log_{10}\left( {M\over \MSun} \right) \\ \nonumber
  &&- 0.09 \log_{10}\left( {M \over \MSun} \right)^2.
  \label{Eq:IMFMmax}
\end{eqnarray}
We performed simulations for those clusters that, upon generating the
initial mass-functions have at least one star of at least
20\,\MSun.

\begin{figure}
\includegraphics[width=0.95\columnwidth]{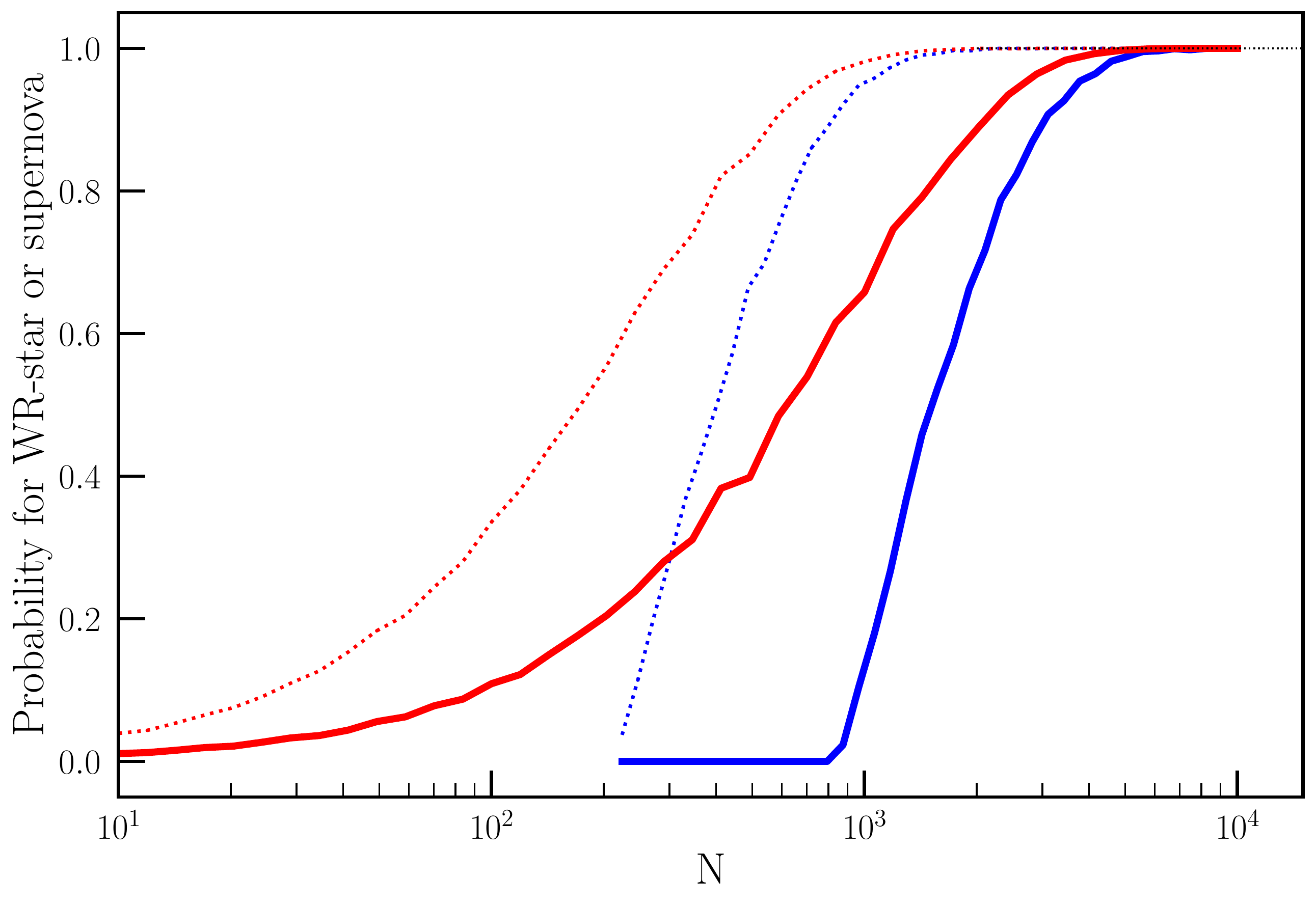}
\caption{Probability density function for acquiring at least one
  Wolf-Rayet star (solid curves) or at least one star sufficiently
  massive to experience a supernova (dotted curves) from the adopted
  broken power-law initial mass-function.  The red curves give the
  probability for an initial mass function with an upper limit of
  120\,\MSun\, the blue curves for an upper-mass cut-off as presented
  in Eq.\,\ref{Eq:IMFMmax}.
  \label{fig:WRpropability}
}
\end{figure}

In Fig.\,\ref{fig:WRpropability} we present the distribution for the
probability of having at least one Wolf-Rayet star or at least one
star sufficiently massive to experience a supernova in a cluster. For
a mass-function with a general upper limit (of 120\,\MSun) independent
of the number of stars in the cluster, there is a finite probability of
finding a Wolf-Rayet star even in a very small cluster of only
10\,stars, whereas for an upper-mass limit according to
Eq.\,\ref{Eq:IMFMmax} such a cluster requires to be composed of at
least some 900 stars. In \S\,\ref{Sect:IMFUpperMassLimit} we discuss
this upper mass limit in some more detail.

Eventually, in the calculation of the relative birth rates
we correct for this bias by multiplying the cluster
probability-density function with the probability of generating such a
massive stars (see \S\,\ref{Sect:Results}). With the adopted mass-function, this probability
entails
\begin{equation}
    P_{m>20\MSun} \simeq 1.0 - {1.2 \over 1 + \left( {N/1350} \right)^{2.70}}.
\end{equation}
We present this distribution in Fig.\,\ref{fig:WRpropability},
including the probability for finding a Wolf-Rayet star in a cluster
with a fixed upper limit of 120\,\MSun\, to the mass function.

\section{Comparison with the Solar system} 

We study the probability that a cluster produces a star with a disk
similar to the Solar system.  For this, we assign a Solar-system
similarity parameter, ${\cal S}_{\rm sp}$, to each star at the end of
each simulation (at $t=50$\,Myr). This parameter is the normalized
phase-space distance in terms of stellar mass and disk size, mass, and
relative inclination. The phase space distance for each of these
parameters are determined by comparing the result of the simulations
with Gaussian distributions around the Solar system characteristics
(for stellar mass $1.0\pm 0.1$\,\MSun, and the disk parameters: radius
$r_{\rm disk} = 45\pm10$\,au, mass $m_{\rm disk} = 0.01\pm
0.003$\,\MSun and relative inclination $i_{\rm disk} =
5.6\pm1.2^\circ$).
\begin{equation}
  {\cal S}_{\rm sp} = S_{\rm stellar-mass} \times S_{\rm disk-mass}
                   \times S_{\rm disk-size} \times S_{\rm inclination}
\end{equation}
For the Solar system ${\cal S}_{\rm sp} \equiv 1$.  The sum of the
values of ${\cal S}_{\rm sp}$ for all the stars in a cluster is the
expectation value for Solar system analogs in a particular star
cluster, $N_{\rm ssa}$.

The abundance of SLR is not part of the definition of ${\cal S}_{\rm
  sp}$, but in practice, it turns out that stars with a high value of
${\cal S}_{\rm sp}$ also have high concentrations in \Al\, and
sometimes also in \Fe\, (see Fig.\,\ref{fig:t_since_enrichment}).  To
further mediate the discussion, we compare the observed abundance in
\Al\, and \Fe\, directly with the results of the simulations. In the
latter, it is conceptually easier to consider absolute abundances, and
therefore we calculate these also for the Solar system.  The observed
Solar system's abundance of \Al/$^{27}$Al$ = 4.5$---$5.2 \times
10^{-5}$~\citep{1995Metic..30..365M,2008LPI....39.1999J}.  The total
amount of \Al\, in the Solar System can be calculated from
\Al/$^{27}$Al $\times$ \Al/$^{1}$H $\times$ Z$_\odot \times
M_{26}/M_{1} = 2.9$---$3.4 \times 10^{-9}$\,\MSun/\MSun\, \citep[see
  also ][]{2015AA...582A..26G}.  Here we adopted the canonical value
of $^{27}$Al$/^1$H$ = 3.5 \times 10^{-6}$ \citep{2003ApJLodders} and
$Z_\odot = 0.71$.  For \Fe\, we use $^{56}$Fe$/^1$H$ = 7.3 \times
10^{-7}$ \citep{2003ApJLodders}, and the observed \Fe/$^{56}$Fe$ = 3.8
\pm 6.9 \times 10^{-8}$ \citep{2018ApJ...857L..15T} which results in
\Fe/$^{56}$Fe $\times$ \Al/$^{1}$H $\times$ Z$_\odot \times
M_{56}/M_{1} = 1.1 \pm 2.0 \times 10^{-12}$\,\MSun/\MSun.

Note that the abundance in \Fe\, derived by
\citep{2018ApJ...857L..15T} is several orders of magnitude lower than
those of \cite[][\Fe/$^{56}$Fe$ \simeq 7 \times
  10^{-7}$.]{2016E&PSL.436...71M}. If the abundances in \Fe\, would
indeed be as high as argued in \cite{2016E&PSL.436...71M}, a supernova
would be unable to enrich the Solar system sufficiently to explain the
current abundance \citep[see
  also][]{2007ApJ...662.1268O,2010ApJ...711..597O}.

\section{Results}
\label{Sect:Results}

\subsection{The most probable birth cluster}

\begin{figure}
\includegraphics[width=0.95\columnwidth]{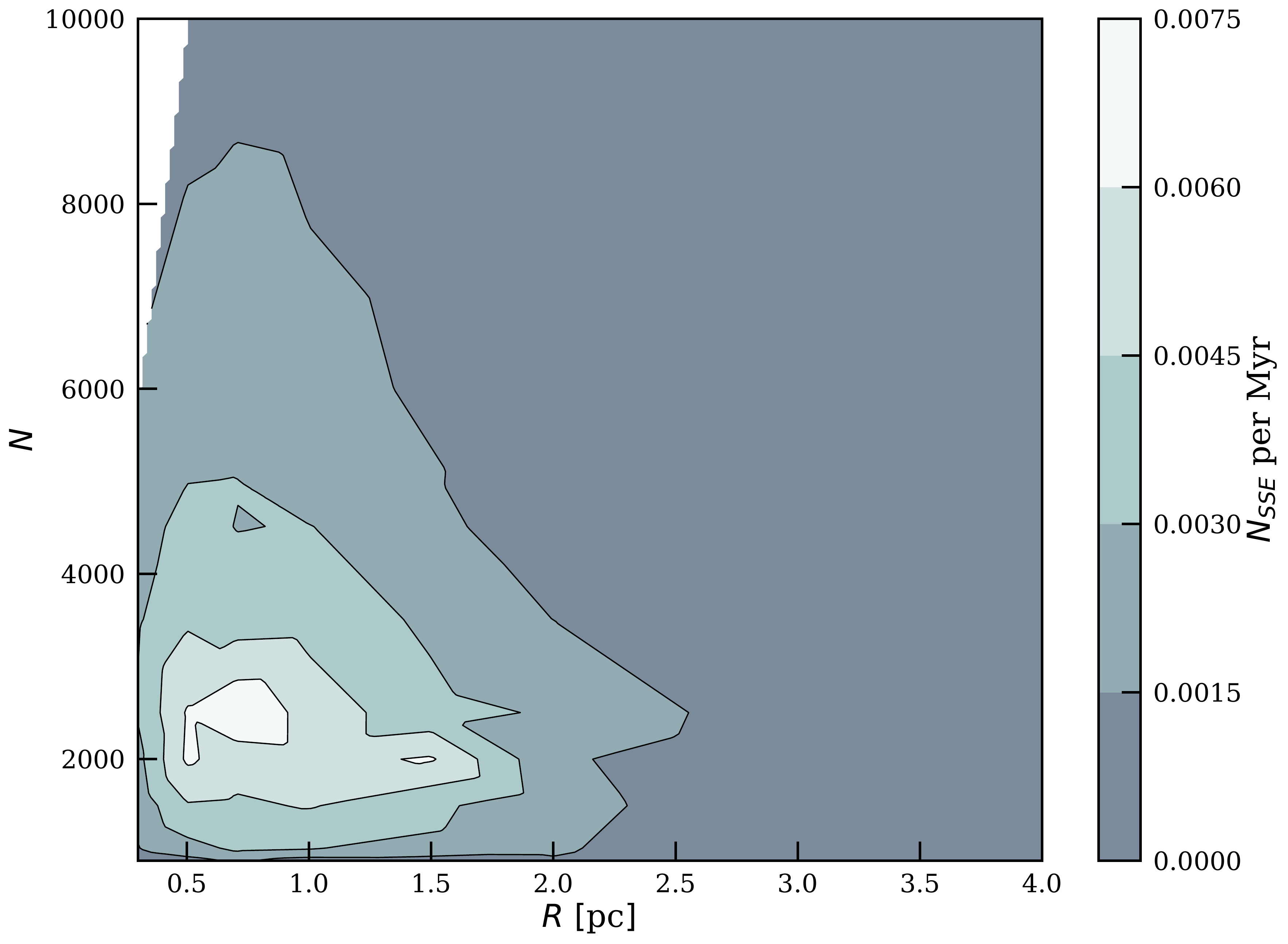}
\caption{Expected number of Solar-system equivalents formed per Myr in
  the Galaxy for clusters with a virialized ($q=1$) Plummer density
  distribution. The calculations are performed in a grid of 10 in mass
  (between 900 stars and $10^4$\,stars) and radius (between 0.1\,pc
  and 4.0\,pc).  The sum of the phase-space distances for each cluster
  is convolved with the star-cluster birth-function in mass
  \cite[using a][function with $\alpha=1$, $\beta=-2.3$ and $m_{\rm
      break}=2\times 10^{5}$\,\MSun]{1976ApJ...203..297S} and radius
  \citep[using a log-normal distribution with $r_{\rm mean} =5$\,pc
    and $\sigma = 3$\,pc,][]{2013ApJ...779..114V}, the probability to
  host a star of at least 20\,M$_\odot$ (see \S\,\ref{Sect:WRWinds}),
  and assuming a star-formation rate of 1\,\MSun/yr
  \citep{2010ApJ...710L..11R}.
  \label{fig:SSBR}
}
\end{figure}

After having performed the grid of calculations, we analyze the
results, in particular with respect to the number of Solar system
analogs $N_{\rm ssa}$.  In Fig.\,\ref{fig:SSBR} we present this
number for the Galaxy (per Myr per cluster) by convolving the
probability density function ($N_{\rm ssa}$ from the results of the
simulations) with the expected number of clusters formed in the Galaxy
(in mass and size) \citep{2010ARA&A..48..431P} and with the observed
Galactic star-formation rate of 1\,\MSun/yr
\citep{2010ApJ...710L..11R}.  The expectation value for the number of
Solar system equivalents born per Myr then is
\begin{equation}
  N_{\rm sse} = N_{\rm ssa} \times f_{\rm Sch}(N) \times f_{\rm ln}(r_{\rm vir}) \times P_{m>20\MSun}.
\end{equation}
Here $f_{\rm Sch}(N)$ is the Schechter \citep{1976ApJ...203..297S}
function with $\alpha=1$, $\beta=-2.3$ and $m_{\rm break}=2\times
10^{5}$\,\MSun) and $f_{\rm ln}(r_{\rm vir})$ is the log-normal
distribution with $r_{\rm mean} =5$\,pc and $\sigma = 3$\,pc)
\citep{2013ApJ...779..114V}.

Massive clusters tend to produce more Solar-system equivalents, but
when integrated over the star-cluster mass function and expected size
distribution the most probable host appears to have a radius of
$r_{\rm vir} \simeq 0.75\pm0.25$\,pc and contains about $2500\pm300$
stars (see fig.\,\ref{fig:SSBR}). This value is consistent with the
earlier derived quantification of the Sun's birth cluster
\citep[][assuming ${\cal S}_{\rm sp} > 0.01$ for each
  parameter]{2009ApJ...696L..13P,2010ARA&A..48...47A,2012MNRAS.419.2448P}.
Clusters with these parameters produce $N_{\rm ssa} = 21.0\pm 5.1$
Solar system equivalents.  In comparison, a cluster with $N=1500$ with
$r_{\rm vir} = 1.0$\,pc produces only $N_{\rm ssa} = 7.0 \pm 2.8$.  We
performed a total of 180 runs with these optimal parameters ($r_{\rm
  vir} \simeq 0.75$\,pc and $N=2500$).

\begin{figure}
\centering
\includegraphics[width=0.95\columnwidth]{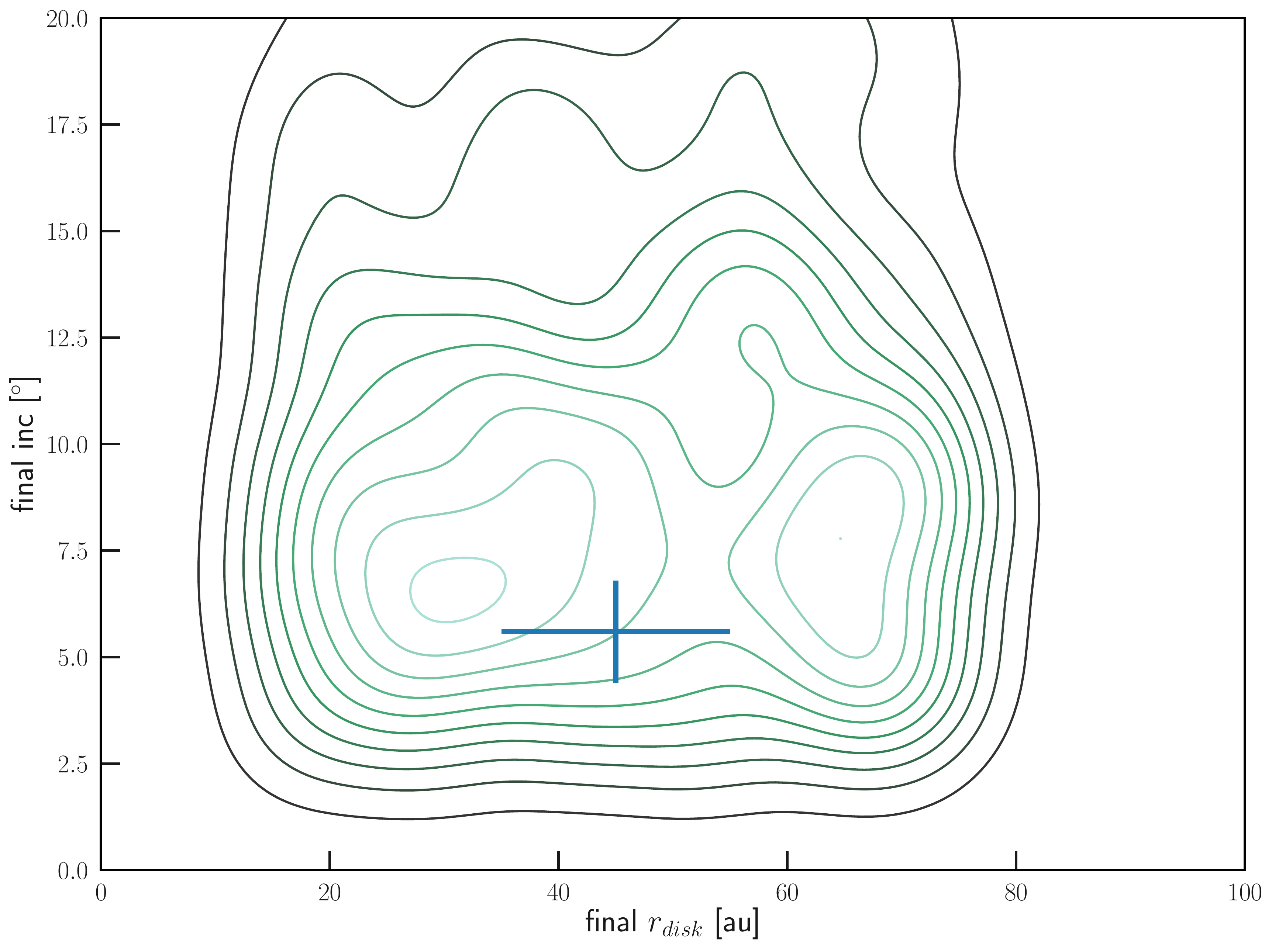}
\caption{Probability density function for the final disk size versus
  relative inclination, for those systems with a high value of ${\cal
    S}_{\rm sp}$.  The error bar indicates the Sun's current parameters.
  \label{fig:disksize_vs_inclination}
}
\end{figure}

In Fig.\,\ref{fig:disksize_vs_inclination} we present the distribution
of disk size and disk inclination at an age of 50\,Myr. The fact that
this distribution is centred around the Sun should not be a surprise,
because both criteria were used to determine a high value of ${\cal
  S}_{\rm sp}$.

\begin{figure}
\includegraphics[width=0.95\columnwidth]{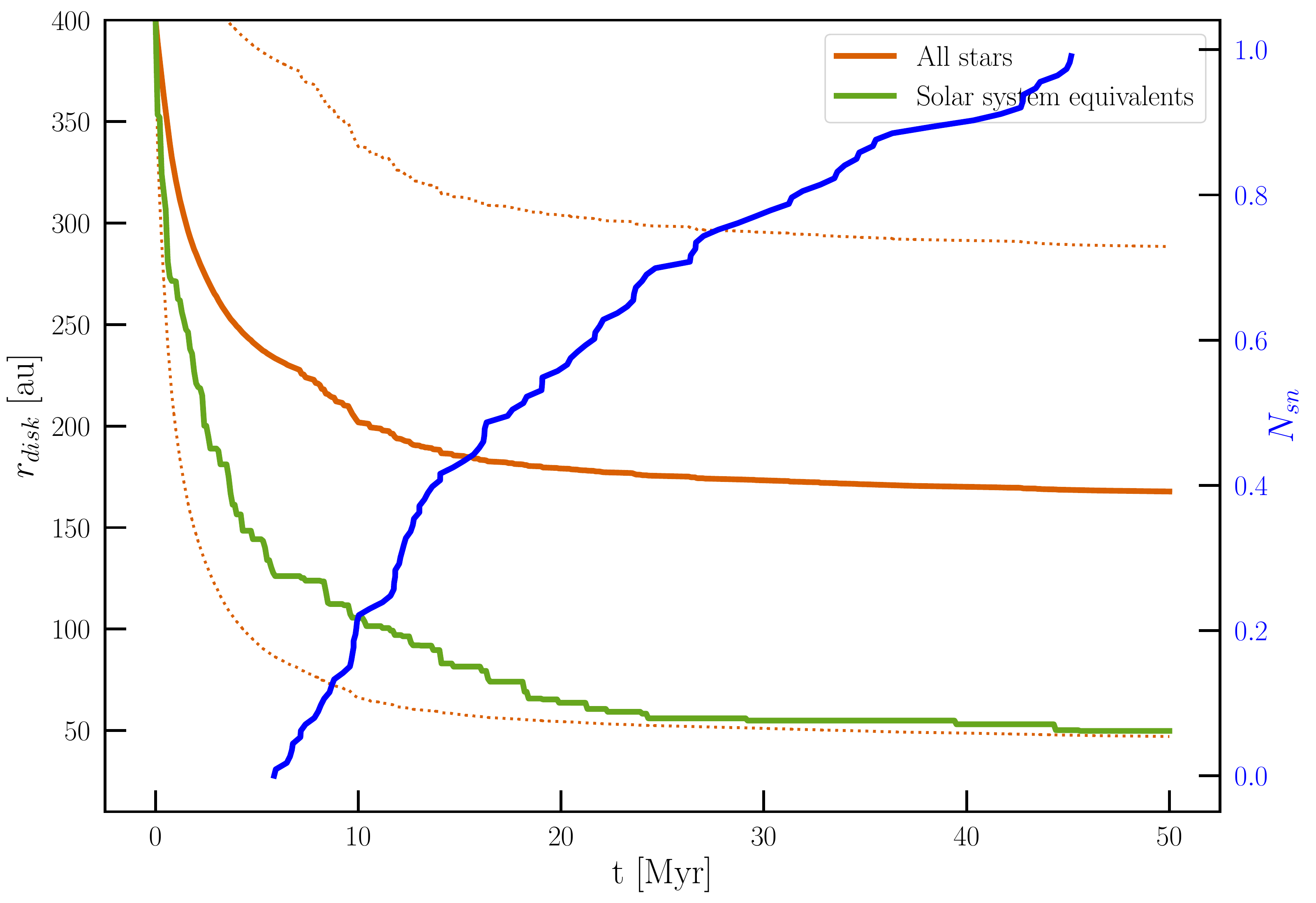}
\caption{Time evolution of the mean disk size (thick orange curve) for
  all 180 simulations with $N=2500$ and $r_{\rm vir} = 0.75$\,pc.  The
    two dotted orange curves give the standard deviation below and above the mean distribution of disk-sizes.  The green curve (below
    the orange curve) gives the evolution of mean disk size for stars
    with a high value of ${\cal S}_{\rm sp}$.  The blue curve (axis to
    the right, also in blue) gives the cumulative distribution of the
    number of supernovae in these simulations.
  \label{fig:disksize_evolution}
}
\end{figure}

In Fig.\,\ref{fig:disksize_evolution} we present the time evolution of
the average disk size, and the cumulative distribution of supernova
occurrences. In the first few ($\sim 6$) Myr the truncation of disks
is driven by close encounters. The introduction of the supernova
truncation process is clearly visible in the mean disk-size
evolution. After about 14\,Myr, supernovae become less dominant again
in terms of disk truncation, and both processes contribute about equal
to the size evolution of the disks. 

\begin{figure}
\centering
\includegraphics[width=0.95\columnwidth]{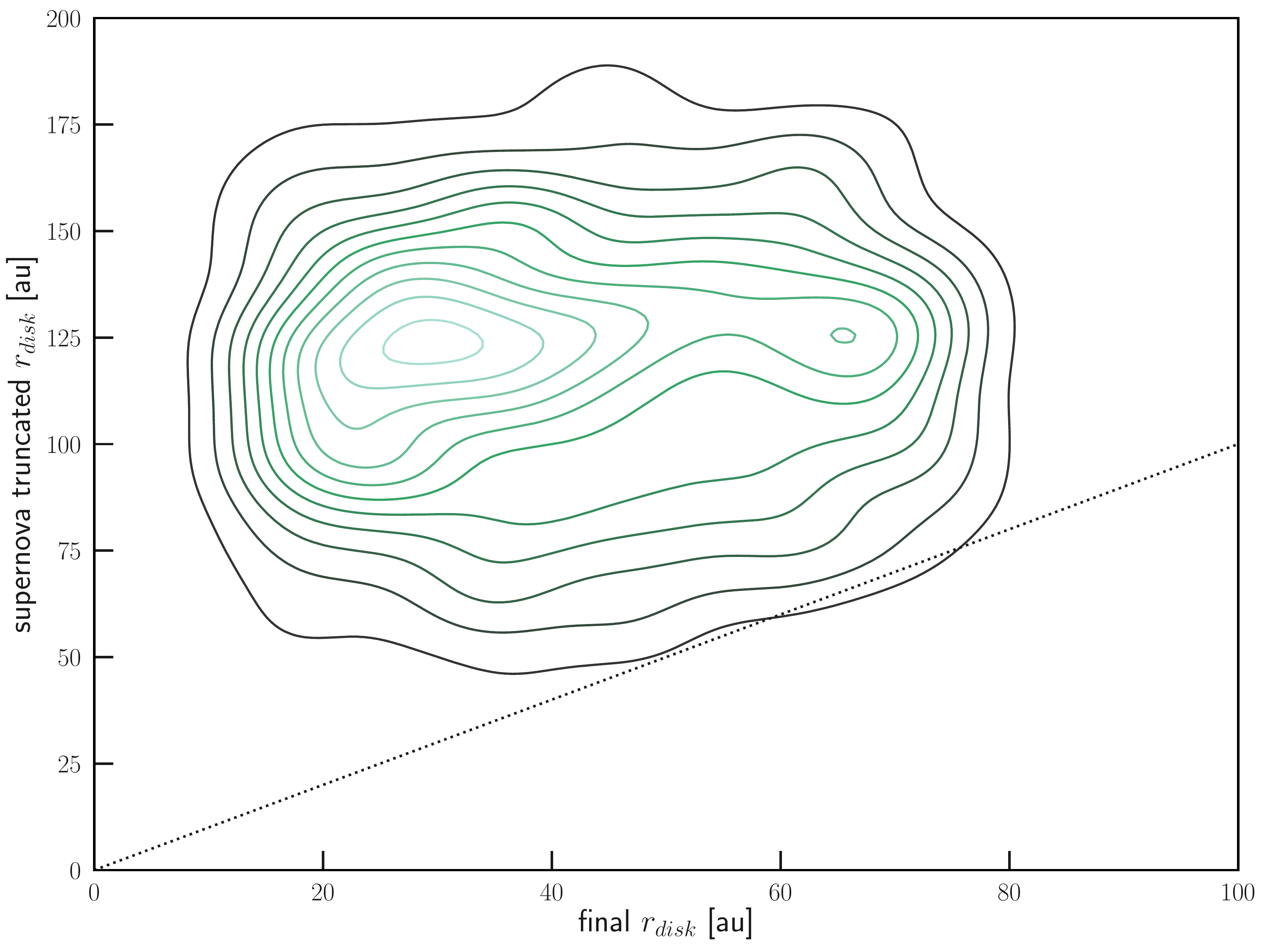}
\caption{The final disk radius (at $50$\,Myr) versus the minimum disk
  radius induced by a nearby supernova for those systems in which the
  disk was most severely truncated by the supernovae.  These account
  for the majority (80\%) of the cases, in the other $\sim 20$\% the
  supernova was more effective in truncating the disk than a nearby
  encounter.  The iso-density contours are given for equal decadal
  percentages. The dotted curve gives the equal radius
  \label{fig:final_vs_sn_disksize}
}
\end{figure}

\begin{figure}
\centering
\includegraphics[width=0.95\columnwidth]{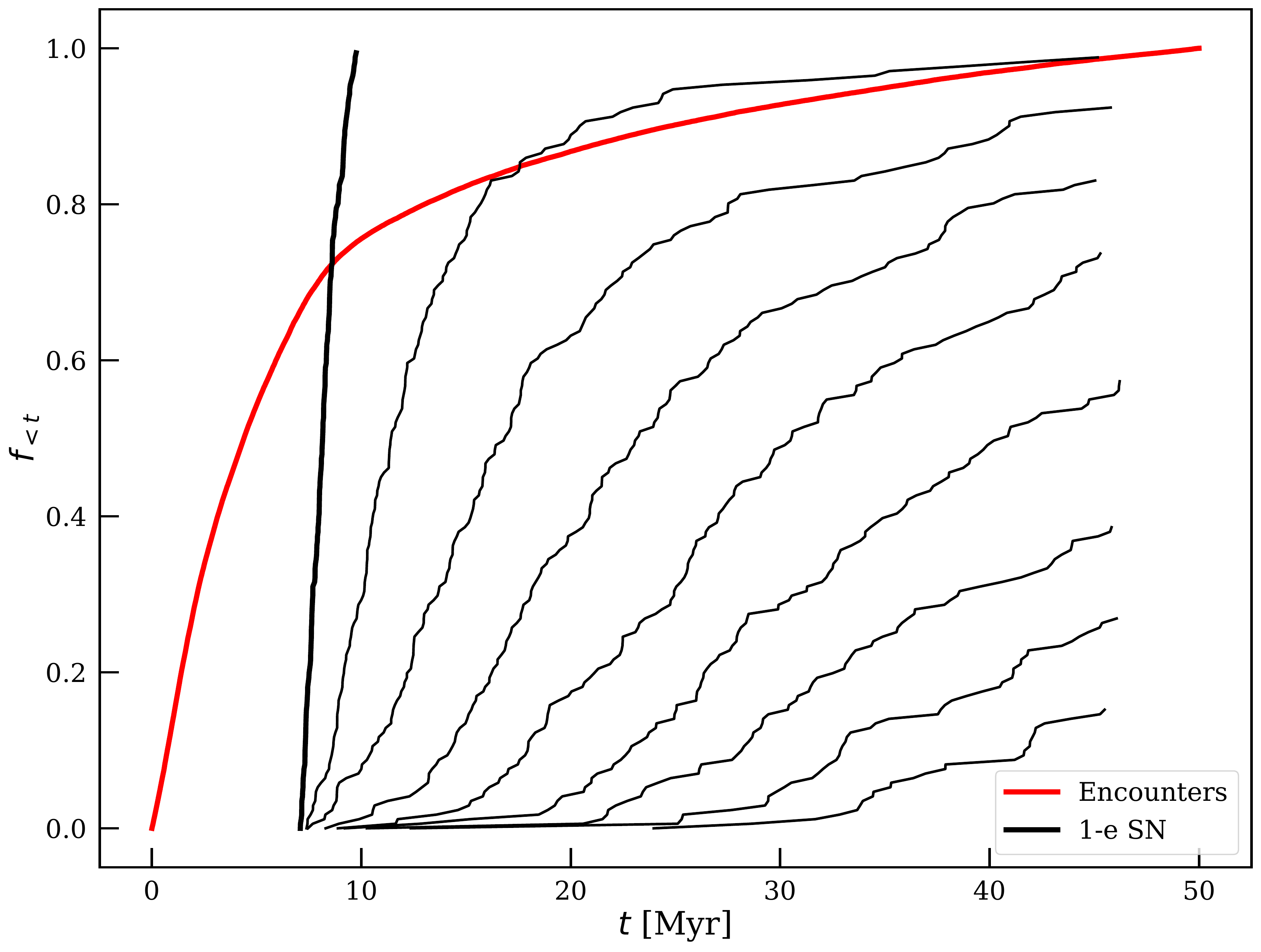}
\caption{Cumulative distributions of the time of the strongest
  encounter (orage) and the moment of the first preserving supernova
  (blue) for solar system equivalent systems. We only plot the
  distributions for the first 9 supernova from the 180 simulations
  with $N=2500$ and $r_{\rm vir} = 0.75$\,pc A sumilar distribution of
  supernova distance was also presented in \cite{2014MNRAS.437..946P}.
  \label{fig:tenc_vs_tsn}
}
\end{figure}

Close stellar encounters are more effective in truncating disks than a
nearby supernova, and they more frequent (in $\sim 80$\% of the cases)
truncate disks to their final value.  In
Fig.\,\ref{fig:final_vs_sn_disksize} we present the distribution of
solar system equivalents for which the disk was more severely
truncated by an encounter than by a nearby supernova.

The majority ($\sim 80$\%) of the protoplanetary disks are severely
truncated by a dynamical encounter before the supernova affects the
disk, but generally close stellar encounters have the greatest impact
on the disk size \cite[see also][]{2015A&A...577A.115V}.  We
illustrate this in Fig.\,\ref{fig:final_vs_sn_disksize}, where we plot
for solar system equivalents the eventual disk radius as a function of
the truncation due to the most constraining supernova.

In Fig.\,\ref{fig:tenc_vs_tsn} we also present the cumulative
distribution of the number of encounters as a function of time; about
60\% of the truncating encounters have already occurred by the time
the first Wolf-Rayet star evolves.  By the time the supernova starts
to preserve the proto-planetary disks, about half those disk have
already experienced their strongest encounter with another star. But
when supernovae effectively peter out after about 20\,Myr, encounters
continue to be important until the end of the simulation.

\begin{figure}
\centering
\includegraphics[width=0.95\columnwidth]{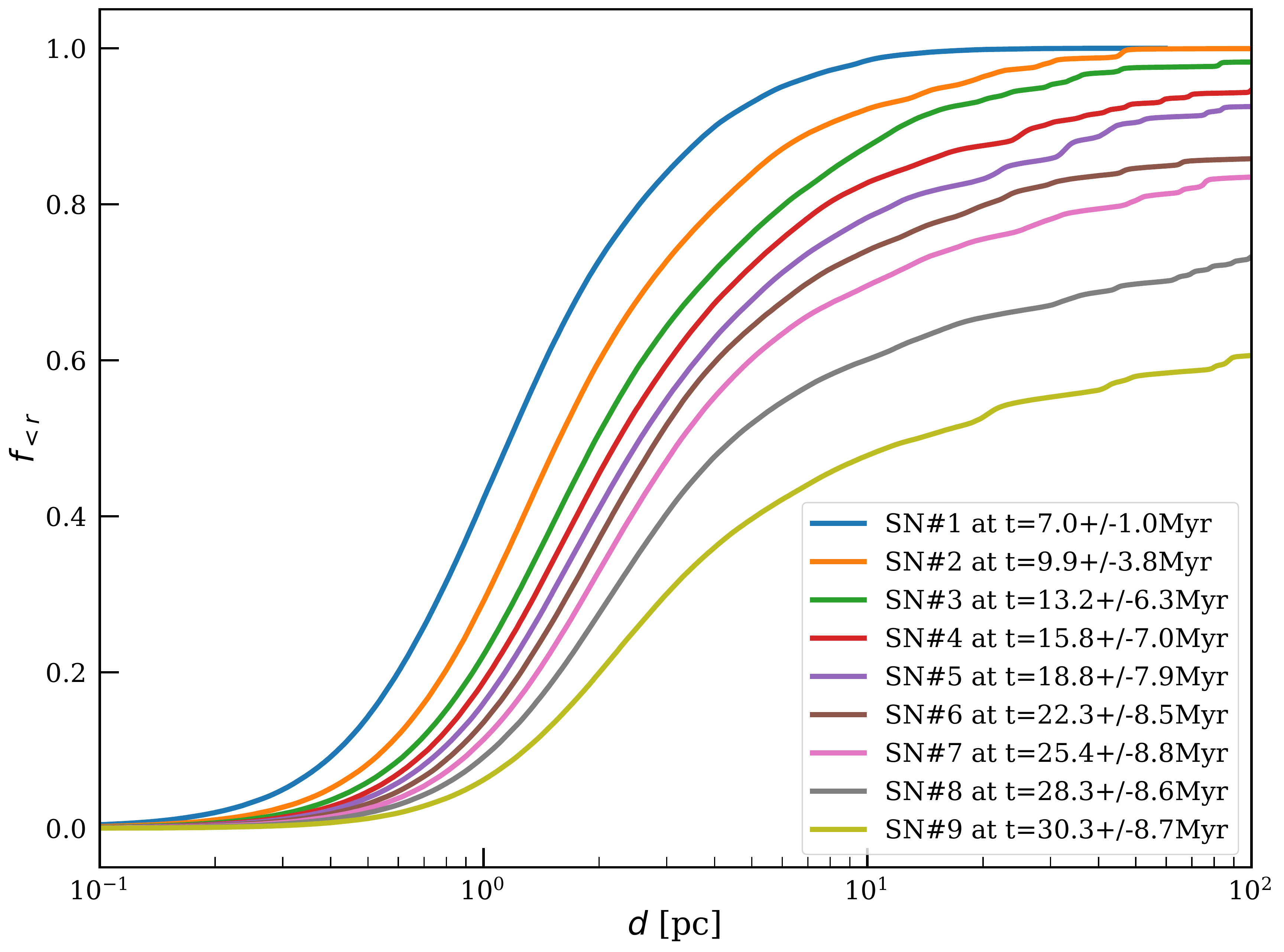}
\caption{Cumulative distributions of the distance between an exploding
  star and the other stars in the cluster, separated in the order in
  which the supernovae occur and normalized to the number of runs.  We
  only show the distance distribution for the first 9 supernovae, even
  though one run experienced as many as 17 supernovae.
  \label{fig:cdf_sn_distance}
}
\end{figure}

We counted as many as 19 supernovae in the clusters that are most
likely to produce a Solar system analogue, but some clusters only
experience 3 supernovae.  In fig.\,\ref{fig:cdf_sn_distance} we
present the cumulative distribution of the distance between a
supernova and the other stars in the cluster, for the first 9
supernovae. We separated the curves for the different supernovae in
the simulation, normalized to the first supernova. In the legend, we
present the moment in time and $1 \sigma$ uncertainty for the various
supernovae. The first supernova typically occurs at an age of
$7.0\pm1.0$\,Myr, and the 9th supernova, which occurs in about 60\% of
the simulations is at an age of $30.3\pm8.7$\,Myr. This figure
indicates that all stars are affected to some degree by at two
supernovae and that in the majority is simulations quite a larger
number of supernovae occur \citep[see also][]{2017MNRAS.464.4318N}.
There is no correlation, however, between the number of supernovae and
the expected number of Solar system equivalents.

\begin{figure}
\includegraphics[width=0.95\columnwidth]{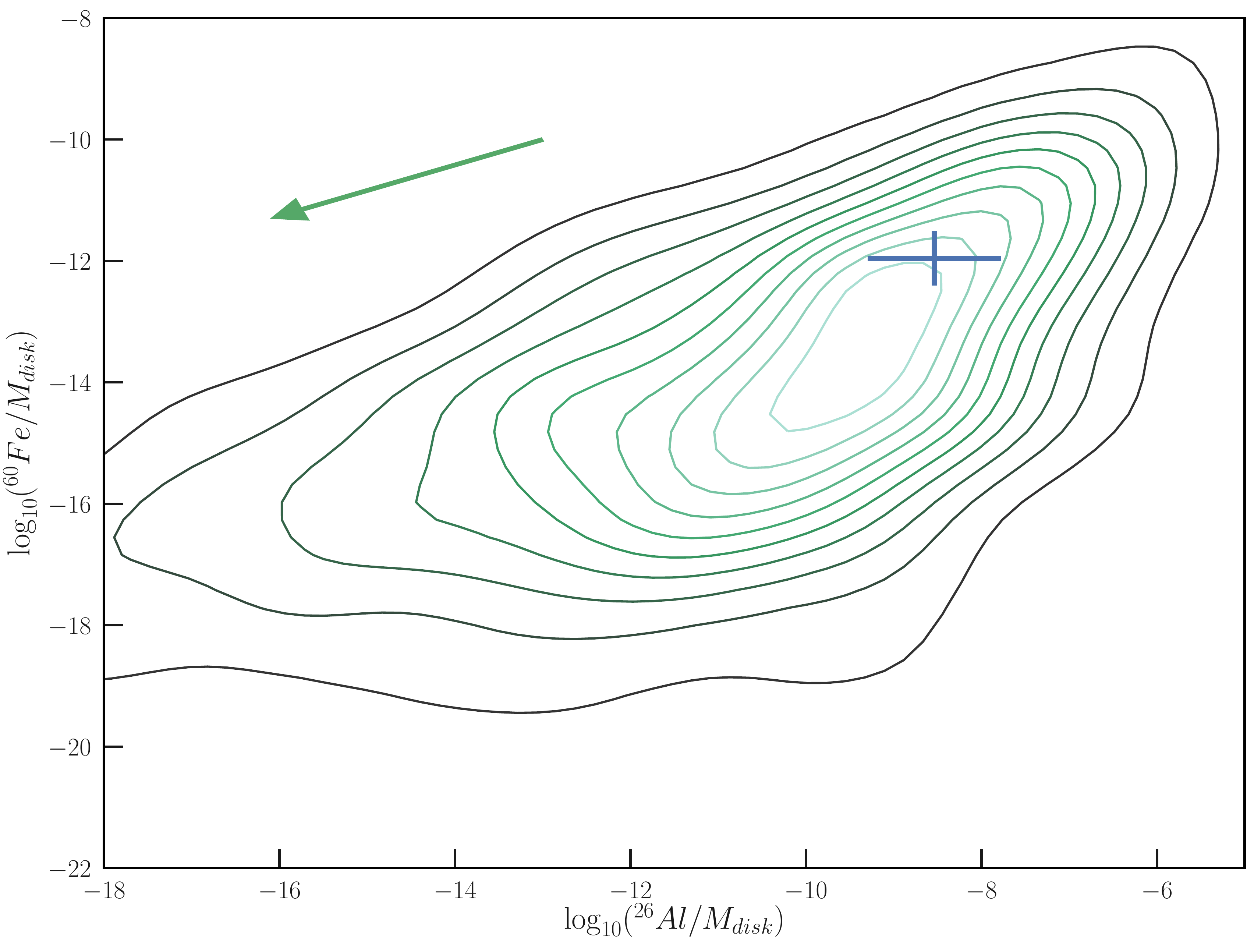}
\caption{Distribution of the preserved abundances in \Al\, and \Fe\,
  of proto-planetary disks for which ${\cal S}_{\rm sp}>0.01$.  The
  equi-composition contours at constant elevation result from the
  superposition of 180 simulations performed with $N=2500$ stars with
  a virial radius of $r_{\rm vir} =0.75$\,pc that resulted in 808
  Solar system equivalents.  The error-bar indicates the abundance of
  the Solar system, with a size equivalent to the uncertainty in the
  measured values.  The green arrow indicates the 10\,Myr change in
  the abundance in \Al\, and \Fe\, due to nuclear decay.  Note that
  the distribution is very broad, even the peak of the distribution
  ranges over more than 2 orders of magnitude in abundance.
  \label{fig:t_since_enrichment}
}
\end{figure}

The distribution of the abundances in \Al\, and \Fe\, for the most
likely birth cluster, presented in Fig.\,\ref{fig:t_since_enrichment},
is consistent with the observed abundances. For other cluster
parameters the value of ${\cal S}_{\rm sp}$ drops and the comparison
with the amounts of accreted SLRs is worse. Sub-virial ($q=0.4$ and
$0.7$) Plummer distributions tend to produce fewer Solar system
equivalents by about a factor of two, but the abundance in \Fe\, is
typically about an order of magnitude higher. Super virial clusters
($q=1.3$) and fractal initial density distributions (using a fractal
dimension of $F=1.6$) tend to underproduce the number of Solar system
equivalents by about an order of magnitude.

In clusters with an initial fractal distribution many close encounters
tend to reduce the disk-size of most of the stars too well below the
observed Solar system disk. These disks subsequently, have too small a
cross-section to effectively accrete material from the Wolf-Rayet
winds or supernova blast wave. Super-virial clusters tend to expand so
rapidly that by the time the Wolf-Rayet stars and supernovae become
effective, the stars have moved away already too far to be strongly
affected by their outflows and irradiation.

\subsection{The story of the three stars}
\label{Sect:RGB}

\begin{figure*}
\includegraphics[width=2.0\columnwidth]{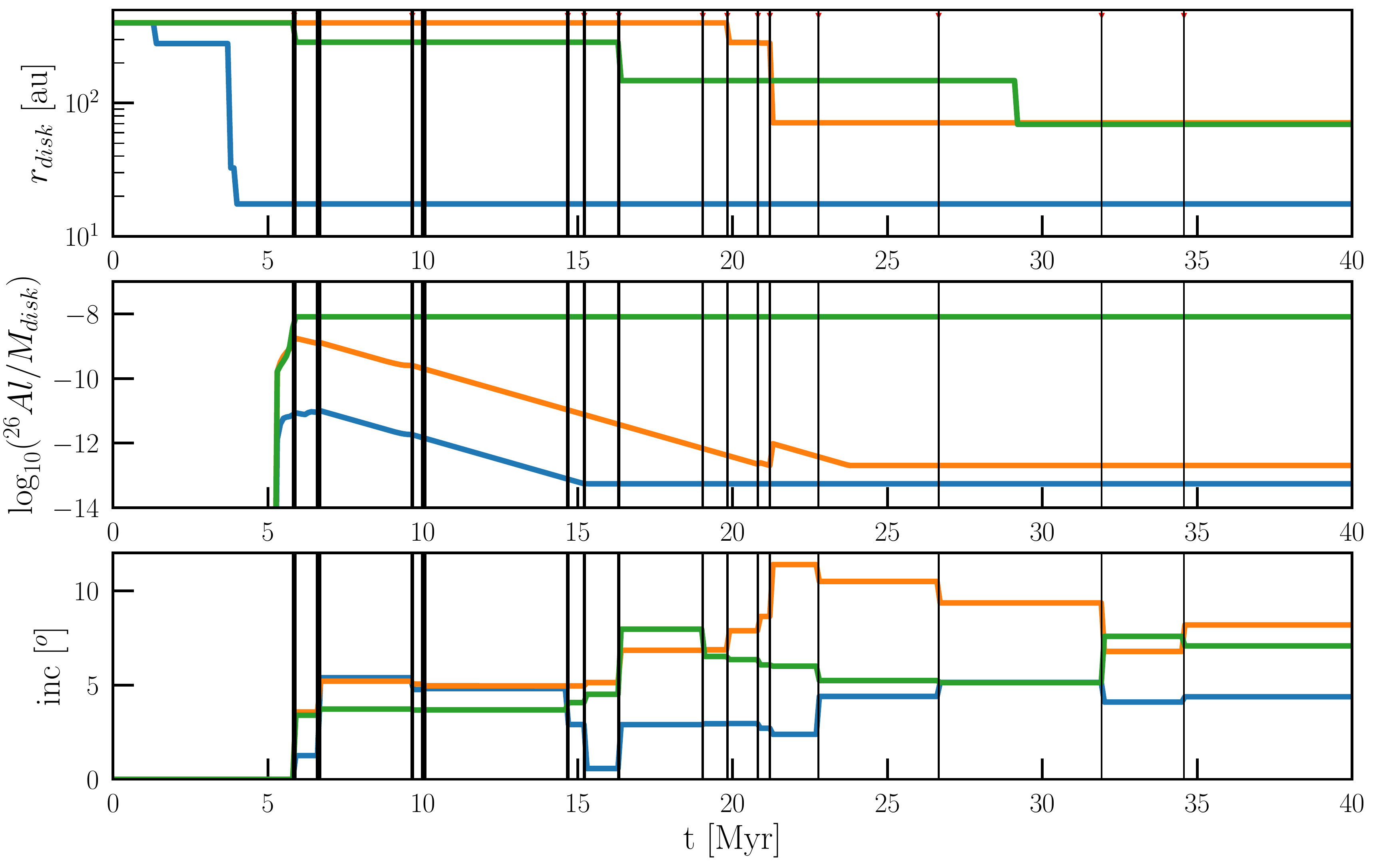}
\caption{Example of the evolution of three (out of seven Solar system
  equivalen) stars in one particular simulations for $N=2500$ and
  $r_{\rm vir} = 0.75$\,pc towards a high value of ${\cal S}_{eqiv}$.
  The coloured lines (green, orange and blue) identify each the
  evolution of one star in time. The vertical lines identify the
  moments of a supernova in the cluster, the thickness of this line is
  proportional to the amount of mass that was injected in the
  supernova shell.
  \label{fig:individual_cluster}
}
\end{figure*}

The evolution of a star that eventually acquires a high value for its
Solar system equivalent is rather typical.  To illustrate this we
present here the story of three stars, which we call green, orange and
blue.  Their evolution is illustrated in
fig.\,\ref{fig:individual_cluster} where we show three panels in disk
size, relative \Al\, abundance of the disk and inclination for 3 out
of 7 stars with the highest value of ${\cal S}_{\rm sp}$ from one
particular simulation with $N=2500$ and $r_{\rm vir} = 0.75$\,pc, of
which we have run 180 realizations.

Upon the birth of the stars, dynamical encounters start to become
effective in truncating the protoplanetary disks after about a
Myr. The first Wolf-Rayet star starts to inject \Al\, enriched
material in the cluster from about 5\,Myr, enriching the
protoplanetary disks in its vicinity. We follow three of the stars
which we eventually classify as viable Solar system analogues. Their
evolutions are presented in Fig.\,\ref{fig:individual_cluster}.

\subsubsection{The story of green}
\label{Sect:Green}

The green star starts like any other with a disk of 400\,au and without
any \Al. Shortly after 5\,Myr the Wolf-Rayet star in the simulation
starts to lose mass, from which the green star accretes quite
effectively resulting in an increase in its relative abundance in
\Al. The explosion of the Wolf-Rayet star does not further affect the
amount of accreted \Al\, but truncates its disk to about 300\,au. The
supernova was sufficiently close to preserve the disk composition,
which remains constant for its remaining lifetime. The next nearby
supernova occurs at an age of about 15\,Myr, which further truncates
the disk, but does not further increase the amount of \Al. Later, at
an age of about 28\,Myr, the star has a close encounter, further
reducing the disk size to about 70\,au. The evolution of the
inclination of the disk represents a random walk and ends at a
relative inclination of about $8.2^\circ$.  Interestingly, by the end
of the simulation, this star has no appreciable amount of \Fe.  The
first supernova preserved the disk and did deposit some small amount
of \Al. The accreted SLRs, however, continued to decay until the 7th
supernova at about 16\,Myr truncated the disk, but it did not deposit
much new \Fe\, and it was too far away to heat the disk appreciably.

The green star was classified as a potential Solar system candidate on
the basis of its stellar mass, disk-size, mass and inclination. The
composition of \Al\, was sufficiently copious to be resembling the
Solar system, but its lack of \Fe\, makes it a less suitable
candidate.

\subsubsection{The story of orange}
\label{Sect:Orange}

The orange star also accretes copiously from the Wolf-Rayet wind, but
when it explodes the star is too far from the supernova to heat or
truncate the disk. The subsequent decay reduces the \Al\, content in
the disk until the short succession of the 10th and 11th supernovae in
the cluster enrich the disk with \Al\, and \Fe, truncate it to about
70\,au and preserve the composition. The disk accretes copiously from
the 11th supernova, but since the irradiation of this supernova
arrived before the \Al\, and \Fe\, enriched blast wave this accreted
material decays again in the following few Myr. Eventually, the
composition settles due to the heating that occurred in the 10th or
11th supernova.  The orange star accreted enormous amounts of \Fe\, to
a level of \Fe/\Al$ = 0.064$, which exceeds the observed value by
about a factor of 42.

\subsubsection{The story of blue}
\label{Sect:Blue}

The blue star experienced multiple rather strong encounters before the
first Wolf-Rayet star started to blow a \Al\, enriched wind. By this
time the disk was already truncated to about 20\,au, and only a small
fraction of the enriched Wolf-Rayet wind is accreted. The first 5
supernovae are not close enough to heat the disk to a temperature
sufficiently high to preserve the SLRs, but eventually the composition
of the disk is preserved at an age of about 15\,Myr.  The blue star
accreted some \Fe\, to a final abundance of \Fe/\Al$ = 8.53 \times
10^{-6}$.

\section{Discussion and conclusions}\label{Sect:Discussion}

We performed simulations of star clusters to study the possible
formation and birth environment of the Solar system. We
envision the early solar system, before the planets formed, but using
the current Solar system parameters as a template. This includes the
mass of the host star, the current size of the disk and the relative
inclination of the ecliptic with respect to the Sun's equatorial
plane. These aspects of the Solar system are prone to external
influences, and they may carry information on the Sun's birth
environment. Equally wise we anticipate that morphological changes to
the young disk perpetuate to later time, and may lead to observable
peculiarities today \citep[see also,][]{2017MNRAS.471.2753R}.

In the calculations, we follow the evolution of these circumstellar
disks, while focussing on the change in disk parameters due to stellar
encounters, the winds of nearby massive stars and the effect of
supernovae. The underlying direct $N$-body code, used to calculate the
dynamical evolution of the cluster in the Galactic tidal field is used
to measure the encounter rate and parameters, but also the distance
and relative velocity of nearby Wolf-Rayet stars and supernovae. The
effects of the latter are used with parametrizations of earlier
radiative-hydrodynamical calculations, which we fitted with respect to
distance and relative inclination angle between the disk orientation
and the line-of-sight to the exploding star. These fit results are
presented in the appendix (\S\,\ref{AppendixA} and
\S\,\ref{AppendixB}).  In the following \S\, we briefly discuss some
of the advances and limitations of this model.

\subsection{The importance of the upper limit to the stellar mass-function
            for a specific cluster mass}\label{Sect:IMFUpperMassLimit}

In the simulations, we adopted an upper limit to the initial mass
function which is a function of cluster mass (see
\S\,\ref{Sect:IMF}). Such a mass limit was already discussed in
\cite{1978stfo.book.....R} and \cite{1982A&A...115...65V}, but
quantified in \citep{2006MNRAS.365.1333W} which, according to
\cite{2015MNRAS.452.1447K}, is the result of the sampling bias and an
underestimate of their error bars \citep{2014PhR...539...49K}.  At
least the Orion star-formation region appears to show evidence for a
deficiency of high-mass stars in low-density environments
\citep{2013ApJ...764..114H}, but it remains unclear if this is a global
phenomenon.

The upper limit of a mass function as a function of cluster mass may
have a profound influence on the dynamical (and chemical) evolution of
star clusters \citep{2014MNRAS.445.2256K}. Also for our study the
consequences are profound. Earlier calculations for studying the
chemical enrichment of the Solar system by \cite{2014MNRAS.437..946P}
study this effect.  When neglecting cluster-mass dependency for the
upper limit to the initial stellar mass function, much lower star
cluster could still host stars sufficiently massive to experience a
supernova or even a Wolf-Rayet star.  \cite{2016MNRAS.462.3979L} and
\cite{2017MNRAS.464.4318N} studied the consequences for the possible
proximity of an exploding star near the proto-solar system in young
(<10\,Myr old) star clusters. They find that the most likely parent
cluster would be about 50-200\,\MSun\, (or about 150 to 600 stars)
with a characteristic radius of about 1\,pc. According to
\cite{2009ApJ...696L..13P}, a proto-solar system would not have the
opportunity to be dynamically truncated but if sufficiently compact
such clusters could still lead to the proper amount of enrichment due
to a nearby supernova, and their close proximity could sufficiently
truncate the disks.  Upon reperforming our simulations, but now with a
fixed upper mass-limit of 120\,\MSun\, for the initial mass-function,
we confirm this result. However, we find that the most optimum cluster
size drops to about 0.2\,pc with a total of about 50 to 100 stars per
cluster, which is somewhat smaller and less massive than found by
\cite{2016MNRAS.462.3979L} and \cite{2017MNRAS.464.4318N}.  With the
fixed upper mass-limit, low mass clusters contribute considerably to
the formation of Solar-system equivalents. The actual contribution of
the number of Solar system equivalents in the Galaxy then depends
quite sensitively on the star-cluster mass function.  How often is a
Solar-mass star born together with a few other stars of which at least
one evolves into a Wolf-Rayet star?

\begin{figure}
\includegraphics[width=0.95\columnwidth]{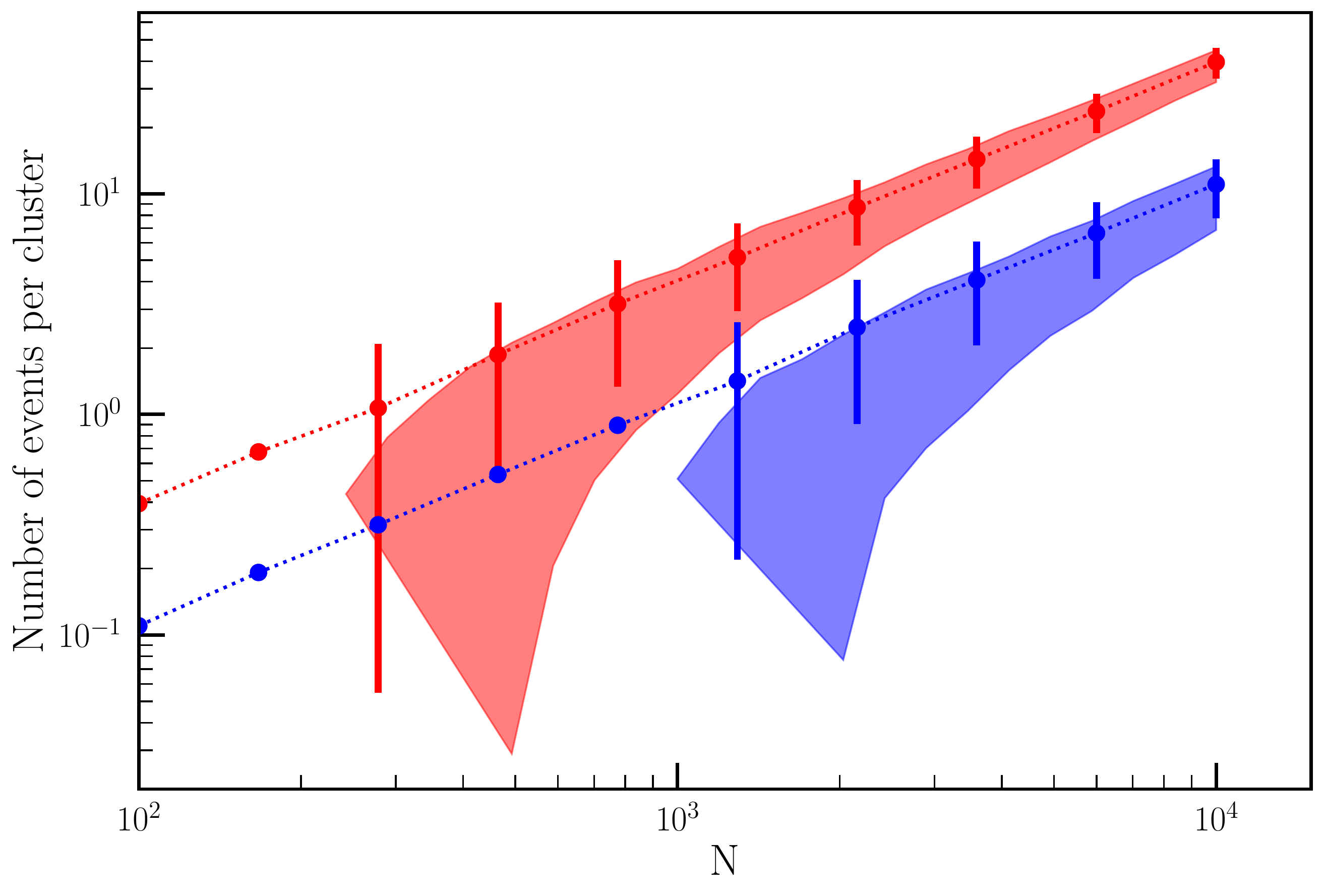}
\caption{Expected number of Wolf-Rayet stars (blue) and supernovae
  (red) in a cluster as function of the number of stars.  The
  red-shaded areas indicate the expectation range (one standard
  deviation around the mean) of stars sufficiently massive to
  experience a supernova for the broken power-law initial
  mass-function with an upper mass-limit as provided by
  Eq.\,\ref{Eq:IMFMmax}.  The blue-shaded area gives the same
  statistics for the number of Wolf-Rayet stars.  The dotted curves
  give the mean of the number of Wolf-Rayet stars (blue) and expected
  number of supernovae (red) for a mass-function with a fixed
  upper-limit of 120\,\MSun.
  \label{fig:SNcount}
}
\end{figure}

Regretfully, we are not going to answer these questions here, but to
help to evaluate the results and possibly scale them to another --lower
mass-- environment in which massive stars are common, we present
fig.\,\ref{fig:SNcount}.  There we present the results of a small
statistical study in which we randomly populate an initial mass
function and count how many stars are sufficiently massive to
experience a supernova (red) or turn into a Wolf-Rayet star (blue).
We perform this experiment with a fixed upper limit to the stellar
mass function of 120\,\MSun\, (dotted curves), and for a mass function
with an upper-mass limit that depends on the cluster mass (see
Eq.\,\ref{Eq:IMFMmax}). The latter are presented as shaded regions.
The largest discrepancy between the curves and the shaded regions are,
not surprisingly, encountered in low-mass clusters.

For the flexible (cluster-mass dependent) upper limit for the mass
function a cluster requires at least $\sim 200$ stars for hosting a
star that is sufficiently massive to experience a supernova, and at
least $\sim 900$\,stars for hosting a Wolf-Rayet star.  With fixed
upper-mass limit of 120\,\MSun\, even a very low-mass cluster can, by
chance, already host such massive stars.

\subsection{The expected disk lifetime and survival}

The processes discussed here to enrich, truncate and tilt
circumstellar disks seem to result in an appreciable number of systems
with characteristics not dissimilar to the Solar system. The
initiation of these processes, however, take a while before they
become effective. A Wolf-Rayet star requires a few Myr, at least, to
start developing an \Al\, rich wind and the first supernova explosion
typically occurs some 8 to 10 \,Myr after the birth of the cluster.
In order to make these processes suitable for affecting the young
Solar system, we require the circumstellar disk still to be present.
This may pose a serious limitation to the proposed chain of events because disks tend to dissolve on a time scale of $\sim 3.5$\,Myr
\citep{2018MNRAS.477.5191R}.  These lifetimes are measured in terms of
timescale on which the disks of half the stars in a particular cluster
drop below the detection limit.  This may indicate that the disk
transforms from a gaseous or dust-rich disk into a more rocky
composition.  The latter is harder to observe with infrared telescopes
and therefore tend to be counted as the absence of a disk.

Other studies argue that disks may survive for much longer.
Statistically, the short lifetimes of disks may be a selection effect
\citep{2014ApJ...793L..34P}, and indeed, there are several environments
which are much older than a few Myr and still have a rich population
of circumstellar disks, such as in the $\eta$\, Chamaeleontis Cluster
with a mean age of $\apgt 10$\,Myr
\citep{2005JKAS...38..241L,2003MNRAS.338..616L}, or the $\apgt
30$\,Myr old Tucana-Horologium Association
\citep{2004ApJ...612..496M}.

At this point, it is unclear if the long time scale poses a problem
for the proposed chain of events.  A method to shorten the time scale
for this model is the introduction of multiple star clusters with
slightly different ages as proposed by \cite{2016MNRAS.456.1066P}, or
a Wolf-Rayet star that runs into the Sun's birth cluster
\citep{2010ApJ...714L..26T}.

The relatively short lifetimes of protoplanetary disks as anticipated
from observations may be the result of the early truncation and the
effects of supernovae, as we discussed here (see also
Fig.\,\ref{fig:disksize_evolution}). We find, in our most probably
parent cluster that the majority of the circumstellar disks are
truncated to about half their initial size on a time scale of about
10\,Myr. In less smooth initial density profiles (such as the fractal
distributions) or more compact clusters, this timescale can be
considerably shorter. In contrast, both clusters for which disk
lifetimes are observed to be long are either low in mass, such as
$\eta$\, Chamaeleontis, or have low density, such as Tucana-Horologium
Association.

\subsection{The effect of encounters and supernova on the later disk morphology}

Most disks in our simulations are severely affected by dynamical
encounters as well as by nearby supernovae. In particular, those disks
that later turn out to acquire parameters not dissimilar to the Solar
system are strongly influenced by both processes. In many cases,
dynamical encounters truncate the disk, after which it is further
abrased and heated by a nearby supernova explosion.  Calculations of
disk harassment have been performed earlier, and they tend to conclude
that stellar encounters have a dramatic effect on the disk-edge
morphology \cite{2014MNRAS.444.2808P,2016ApJ...828...48V}, and that
these effects may be quite common in young star clusters
\cite{2015A&A...577A.115V}. Both arguments are consistent with our
findings.  It is not so clear however, how to quantize the long-term
consequences for a potential Solar system. Some of the induced fringes
on the circum-stellar disk may be lost in subsequent encounters or
supernovae, but other effects may persist and remain recognizable also
at later time \citep{2018MNRAS.tmp.1696R}. A subsequent study,
focussing on the long-term consequences of early truncation of
supernova harassment of the disk would be quite interesting, but this
was not our focus.  As we discussed in \S\,\ref{Sect:RGB} in relation
to Fig.\,\ref{fig:individual_cluster}, disks with similarities to the
Sun's tend to be truncated by dynamical encounters at an early epoch,
and later again by the blast wave of a nearby supernova explosion.
The former mediates the small size of the currently observed disk, but
the latter is important to explain the observed tilt and the accretion
and preservation of SLRs, effectively freezing the disk's composition
in SLRs.  Recently \cite{2018MNRAS.474.5114C} argued that the
misalignment angle between a planetary system and the equatorial plane
of the host star can be explained by dynamical encounters between the
planetary systems in a relatively low-density star cluster. Such
interaction also naturally leads to some of the processed described
here, such as the truncation of the disk.

\subsection{The parental star cluster}

A planetary system typically experiences more than one strong
encounter and more than one nearby supernova. In particular the
latter is Mondial to all stars in the same parental cluster. We
therefore expect many stars to have composition in SLRs not dissimilar
to the Solar system, as well as truncated and inclined disks.

The star clusters which turn out to be most favourable for producing
Solar system equivalents are sufficiently massive that there will be
multiple stars which are sufficiently massive to experience a
supernova. The typical number of supernovae is 9--15, but some
clusters experience as few as 3 and some as many as 19. The first
supernova tends to explode at an age of about 8\,Myr. This is somewhat
constructed because our clusters are initialized with the guarantee of
a $>20$\,\MSun\, star, which for Solar composition has a lifetime of
$\aplt 9$\,Myr and all stars are born at the same time.  In
Fig.\,\ref{fig:tenc_vs_tsn} we present the cumulative distributions
for the time of the first conserving supernova.  The time between the
first and the second supernova ranges between less than 1\,Myr to as
long as 30\,Myr. A short timescale between supernovae is essential in
order to preserve \Fe\, in enriched disks (see
Fig.\,\ref{fig:individual_cluster}). With a half-life of about
2.62\,Myr for \Fe\, and even shorter for \Al, much of the SLRs will
have decayed if the second supernova takes too long. The only clusters
which produce a substantial number of Solar system equivalents
experience two or more supernovae in short succession.

Interestingly, stars born in a Plummer distribution with a high value
of ${\cal S}_{\rm sp}$ tend to produce a wide range of abundances in
\Al\, and \Fe\, with a reasonable match with the observed abundance in
the Solar system (see fig.\,\ref{fig:t_since_enrichment}).  Fractal
clusters fail to reproduce the \Al\, and \Fe\, abundances by several
orders of magnitude. The failure of fractally distributed initial
conditions to produce Solar system equivalents are attributed to the
high degree of dynamical activity in these clusters.  As a
consequence, close stellar encounters in the first few million years
tend to obliterate proto-planetary disks, which subsequently have too
small a cross section to accrete substantial amounts of \Al\, from
Wolf-Rayet winds and \Fe\, from the supernova blast-wave.  this may
change when we take the viscous evolution of the disks into account
(Concha et al., in preparation), in which case truncated disks may
grow back in time.

The cluster that are most likely for form a Solar system analogs
contain $N = 2500\pm300$\,stars ($\sim 900$\,\MSun) with a virial
radius $r_{\rm vir} = 0.75\pm0.25$\,pc, but comparable values are
obtained for somewhat lower mass clusters ($\sim 550$\,\MSun) the
cluster radius can be as large as $r_{\rm vir} \simeq 1.5$\,pc.
Increasing the virial ratio causes a dramatic drop in the expectation
value of ${\cal S}_{\rm sp}$, and cool initial conditions (with a
virial ratio of 0.4 to 0.7) tend to result in a somewhat smaller value
of ${\cal S}_{\rm sp}$, but still with a distinct peak for clusters of
$N\sim 1000$\,stars and $r_{\rm vir} \simeq 0.5$\,pc.  Clusters for
which the stars are initially distributed in a fractal (with dimension
$F=1.6$) produce fewer Solar system equivalents by about a factor of
two compared to Plummer distributions. In these cases, the cluster
that produces most Solar system equivalents is somewhat less massive
($\sim 1000$\,stars) but considerably larger, with a virial radius of
2--3\,pc.

\section{Summary}

Integrating over the phase space in terms of cluster mass and size
results in a Galactic birth-rate of $\sim 30$ Solar system equivalents
per Myr. With an expected lifespan of $\sim 12$\,Gyr we expect the
Galaxy to contains some 36,000 systems with a host-mass, disc-size,
inclination angle, and with abundances in \Al\, and \Fe\, similar to
the Solar system.

We argue that the Sun was born in a cluster of $N = 2500\pm300$\,stars
($\sim 900$\,\MSun) distributed in a smooth potential near virial
equilibrium with a characteristic radius $r_{\rm vir} =
0.75\pm0.25$\,pc. Such clusters produce about 25 planetary systems
with characteristics similar to the Solar system. They have an
abundance in \Al\, and \Fe\, within about two orders of magnitude of
that of the current Sun's planetesimal system, the disk is truncated
to about 45\,au and it is inclined with respect to the star's
equatorial plane.

The evolution of these solar system analogues appears to be rather
typical. The first few Myr are characterized by repeated close
encounters with other stars.  This process causes the disk to be
truncated to a radius of $\aplt 250$\,au\, in average, or to $\aplt
100$\,au\, for systems that are more similar to the Solar system. A
nearby Wolf-Rayet star will enrich the surviving disk with \Al\,
isotopes, which are fried into the disk vitreous droplets upon the
first nearby supernova that is able to heat the disk to $\apgt
1500$\,K. The close proximity of this supernova again harasses the
disk and also injects short-lived \Fe\, isotopes in the disk. This
second generation of enrichment is again fried into vitreous droplets
upon a subsequent nearby supernova. The entire process then takes at
least two supernovae in order to explain the abundances in \Al\, as
well as in \Fe. In principle, a single supernova can be sufficient, but
in that case, the \Fe\, abundance has to be primordial.

\section*{Acknowledgments}
It is a pleasure to thank the referee 
for useful
comments, Arjen van Elteren, Maxwell Cai and Francisca Ramirez-Concha,
and MSc students Fedde Fagginger and Tom Sweegers for discussions.  I
also thank Norm Murray and CITA for the hospitality during a long-term
visit.

\appendix

\section{Boundary conditions and model parameters}
\label{AppendixA}

\subsection{the Astrophysical Multipurpose Software Environment}\label{Sect:AMUSE}

The coupling of the various numerical methods is realized using the
Astrophysical Multipurpose Software Environment ({\tt AMUSE})
\cite{PortegiesZwart2013456,AMUSE}.  {\tt AMUSE} provides a
homogeneous interface to a wide variety of packages which enables the
study of astrophysical phenomena where complex interactions occur
between different physical domains, such as stellar evolution,
gravitational dynamics, hydrodynamics and radiative transport.  For
this project we focus on the coupling between gravitational-dynamics
solvers, stellar-, hydrodynamical and radiative transfer solvers.  The
major advantage of {\tt AMUSE} over other methods is the flexibility
in which a wide variety of solvers can be combined to address the
intricate interactions in the nonlinear multi-scale systems over all
relevant scales.  Also relevant is the way in which we can expand
methods by incorporating additional effects. In this paper, these
effects include the ablation of the circum-stellar disk by a nearby
supernova, the accretion of the winds of nearby stars and the effects
of close stellar encounters.

\subsection{Background Galactic tidal field}\label{Sect:TidalField}

Each cluster is initialized in a circular orbit at a distance of
8.5\,kpc from the Galactic centre.  Since we mainly study relatively
young star clusters we ignore the global evolution of the Galaxy but
assume it to be a slowly varying potential with contributions from the
bar, bulge, spiral arms, disk, and halo. We adopt the same Galactic
parameter as those used in
\cite{2016MNRAS.457.1062M,2017MNRAS.464.2290M} to study the spatial
distribution of Solar siblings in the Galaxy.

\begin{figure}
\includegraphics[width=0.95\columnwidth]{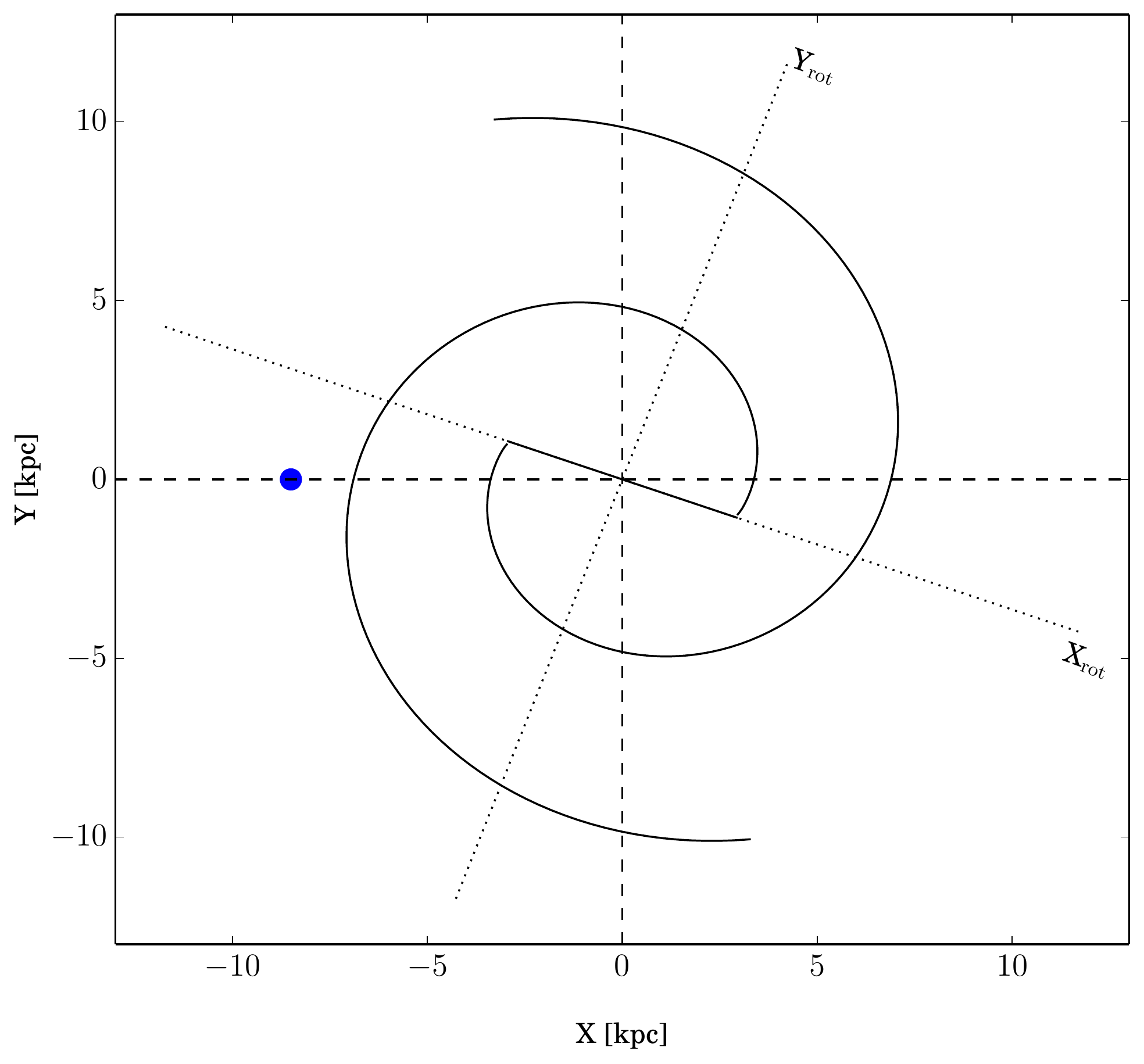}
\caption{Schematic view of the the bar and spiral arms of the Galaxy
  at the present time. The blue bullet marks the current position of
  the Sun measured in an inertial system that is fixed at the centre
  of the Galaxy. The axes X$_\mathrm{rot}$ and Y$_\mathrm{rot}$
  correspond to a system that corotates with the bar. Note that the
  spiral arms start at the edges of the bar, and the coordinates
  (X$_\mathrm{rot_1}, $Y$_\mathrm{rot_1}$) and (X$_\mathrm{rot}$,
  Y$_\mathrm{rot}$) overlap at the present.
  \label{fig:SunInGalaxy}
}
\end{figure}

\begin{table}
  \caption{Modeling parameters of the Milky Way.}
  \label{tab:params}
  \begin{tabular}{lll} \hline
    \multicolumn{3}{c}{ \rule{0pt}{3ex}\textit{Axisymmetric component}} \\ 
    \rule{0pt}{4ex}Mass of the bulge ($M_\mathrm{b}$) & $1.41\times 10^{10}$ M$_{\odot}$ &\\ 
    Scale length bulge ($b_\mathrm{1}$) & $0.3873$ kpc &\\
    Disk mass ($M_\mathrm{d}$) & $8.56\times10^{10}$ M$_{\odot}$  &\\
    Scale length 	1 disk ($a_\mathrm{2}$) & $5.31$ kpc  & 1) \\
    Scale length 2 disk ($b_\mathrm{2}$) & $0.25$ kpc   &\\
    Halo mass ($M_\mathrm{h}$) & $1.07\times 10^{11} $ M$_{\odot}$   &\\
    Scale length halo ($a_\mathrm{3}$) & 12 kpc  &\\ 
    \multicolumn{3}{c}{ \rule{0pt}{3ex}\textit{Central Bar}}   \\ 
     \rule{0pt}{4ex}Pattern speed ($\Omega_\mathrm{bar}$) & $55$ \patspeed & 2)\\ 
    Mass ($M_\mathrm{bar}$) & $9.8\times10^9$ M$_{\odot}$& 4) \\ 
    Semi-major axis ($a$) & $3.1$ kpc & 5)\\
    Axis ratio ($b/a$) & $0.37$& 5) \\
    Vertical axis ($c$) & 1 kpc & 6)\\
    Present-day orientation & $20^\circ$ & 3)\\ 
    \multicolumn{2}{c}{ \rule{0pt}{3ex}\textit{ Spiral arms}} \\ 
     \rule{0pt}{4ex}Pattern speed ($\Omega_\mathrm{sp}$) & $25$ \patspeed & 2)\\
    Number of spiral arms ($m$) & $2$ & 7)\\
    Amplitude ($A_\mathrm{sp}$) &  $3.9\times10^7$ M$_\odot$~kpc$^{-3}$ & 4) \\
    Pitch angle ($i$) & $ 15.5^\circ$ & 4)\\
    Scale length ($R_\mathrm{{\Sigma}}$) & $2.6$ kpc & 7)\\
    Scale height ($H$) & 0.3 kpc & 7)\\
    Present-day orientation & $20 ^\circ$ & 5) \\
    		        \hline
  \end{tabular}\\
  \textbf{References:} 1)
  \citep{1973asqu.book.....A}; 2)  \citep{2011MSAIS..18..185G}; \\
  3) \citep{2011MNRAS.418.1176R}; 4) \citep{2012A&A...541A..64J}; \\
  5) \citep{2015MNRAS.446..823M}; 6) \citep{2014A&A...569A..69M}; \\
    7) \citep{2000A&A...358L..13D}; 8) \citep{2008ApJ...673..864J}
\end{table} 

\subsection{The effects of stellar evolution, mass loss and supernovae}

Stellar evolution is taken into account using the {\tt SeBa} stellar
and binary-evolution code
\citep{1996A&A...309..179P,1998A&A...332..173P,2012A&A...546A..70T,2012ascl.soft01003P}.

Here we use the event-driven time-stepping scheme
between the stellar evolution and the gravitational dynamics codes to
assure that the stellar positions are consistent with the moment a
star explodes in a supernova \citep[see][for details]{AMUSE}.

\subsection{Integrating the equations of motion of the stars in the cluster}
\label{Sect:Nbody}

The equations of motion of the cluster stars are integrated using the
{\tt ph4} 4th-order Hermite predictor-corrector direct $N$-body code
\citep[for implementation details see][]{AMUSE}, with a time-step
parameter $\eta = 0.01$ and a softening of 100\,au.  The Galaxy model
(see \S\,\ref{Sect:TidalField}), is coupled to the cluster dynamics
via the 6th order {\tt rotating-Bridge} method \citep{AMUSE}. This is
a special high-order derived version of the classic bridge
\citep{2007PASJ...59.1095F} but for a rotating reference frame
\citep{2017MNRAS.464.2290M}.  After every 0.1\,Myr (equivalent to the
bridge-timestep) we synchronize the gravity solver, check for energy
conservation, and print diagnostics.  The energy of the $N$-body
integrator is preserved better than $1/10^8$, which is sufficient to
warrant a reliable result \citep{2041-8205-785-1-L3}.

While integrating the equations of motion, we check for close
encounters.  When two stars approach each other in a pre-determined
encounter radius (initially 0.02\,pc) the $N$-body integrator is
interrupted and the system is synchronized.  Such interrupt is called
a {\em stopping condition} in AMUSE lingo and should be perceived as
a numerical trick to stop one code in order to allow another code to
resolve a particular part of the physics.  The encounter is
subsequently resolved and the effect on the disks of the two stars
calculated (see \S\,\ref{Sect:DiskRadius}).

We also keep track of the mass lost by massive $\apgt20$\,\MSun\,
stars during integration.  This mass can be accreted by the circumstellar disks of nearby stars. As a consequence, these disks can be
enriched with the SLR-rich material in the wind. This can lead to the
accretion of \Al\, from nearby Wolf-Rayet stars (see
\S\,\ref{Sect:WRWinds}).

When a star explodes in a supernova, we generate an interrupt and
subsequently, calculate the effects of the irradiation and impacting
SLR-loaded blastwave on all the other stars in the cluster.
Short-lived radionuclides accreted by any of the disks (either by a
supernova or a Wolf-Rayet star) decay with time.  The half-life time
for \Al\, is $\sim 1.1$\,Myr, and for \Fe\, this is $\sim 2.62$\,Myr
\citep{PhysRevLett.103.072502}.  As soon as a supernova heats the disk
to a temperature of at least 1500\,K, we freeze the abundance of SLRs.
At such high temperatures, the intense radiation of the supernova melts
the disk and encapsulates SLRs in vitreous droplets
\cite[see][]{2018arXiv180204360P}.  Subsequent accretion may further
increase the amount of SLRs, which will also decay until another
supernova heats the disk again to a sufficiently high
temperature. This may lead to multiple generations of enrichment. We
apply the same procedure to \Fe\, that is accreted from a supernova
shell.

\subsubsection{The effect of encounters on disk size}\label{Sect:DiskRadius}

The effect of a two-body encounter on the disks of both stars is
solved semi-analytically.  Once a two-body encounter is detected we
calculate the pericenter distance, $p$, by solving Kepler's equation
\citep[using the Kepler-module from the {\tt Starlab}
  package,][]{2001MNRAS.321..199P}.  Note that the closest approach
may be well within the adopted softening radius of 100\,au.  The new
disk-radius for a star with mass $m$ is calculated using
\begin{equation}
  r^\prime_{\rm disk} = 0.28 p \left( {m \over M} \right)^{0.32},
\end{equation}
which was calibrated for parabolic co-panar prograde encounters
\citep{2014A&A...565A.130B,2016MNRAS.457.4218J}. Here $M$ is the mass
of the other star.
This equation is also applied for calculating the new disk radius of
the encountering star.  These new radii are adopted only if they are
smaller than the pre-encounter disk radii.

In order to reduce the number of disk truncations at runtime, and
therewith the number of interrupts (and synchronizations) in the
$N$-body integration, the new encounter distance for both stars is
reset to half the pericenter distance $p$.  This prevents two stars
from being detected at every integration time step while approaching
pericenter, which would cause the disk to be affected repeatedly
during a single encounter.  This procedure, therefore, limits the number
of encounters to the most destructive one at pericentre.

\subsubsection{The effect of encounters on disk mass}

The truncated disks of the encountering stars lose mass.  We estimate
the amount of mass lost from each disk using
\begin{equation}
    \mathrm{d}m = m_{\rm disk} 
           {r_{\rm disk}^{1/2}-r^{\prime 1/2}_{\rm disk} \over r_{\rm disk}^{1/2}}.
\end{equation}
Both encountering stars may accrete some of the material lost from the
other star's disk, which we calculate with
\begin{equation}
  \mathrm{d}m_{\rm acc} = \mathrm{d}m f {m \over M+m}.
\label{Eq:dmdisk}\end{equation}
Here $f \leq 1$ is a mass transfer efficiency factor (we adopted
$f=1$).  Both equations are applied symmetrically in the two-body
encounter, and as a consequence, both stars lose some mass and gain
some of what the other has lost.  Disks may become enriched by
accreting material from earlier enriched disks in close
encounters. Here we assumed that accreted material has the same
composition as the mean host disk.

\subsection{Alumunum-26 enrichment due to Wolf-Rayet winds}\label{Sect:WRWinds}

The copious mass-loss of a Wolf-Rayet star is enriched in \Al\, and
other short-lived radionuclides, but poor in \Fe.  During the
integration of the equations of motion of the stars in the cluster, we
use the stellar evolution code to determine the amount of \Al\,
liberated by the Wolf-Rayet stars.  The amount of mass in the
Wolf-Rayet star is calculated from rotating stellar evolution models
\citep{2004NewAR..48....7V}. We fitted the yields for Solar composition in their Fig.2 to 
\begin{equation}
  \log_{10}(m_{Al}/\MSun) = 1.70\cdot 10^{-8} \left( m/\MSun \right )^{2.29}.
  \label{Eq:WRyield}
\end{equation}
We do not take the time-dependency of the \Al-yields into account, but
adopt a constant mass fraction in the wind assuming that the entire
stellar envelope is homogeneously enriched with \Al\, at the amount
given by Eq.\,\ref{Eq:WRyield}. The stellar-evolution code provides
the appropriate mass-loss rate and wind-velocity.

During the integration of the equations of motion, mass lost by any of
the Wolf-Rayet stars is accreted by the proto-planetary disks of the
other stars. The amount of mass accreted from the wind is calculated
using the Bondi-Hoyle-Littleton accretion \citep{bondy_hoyle44}
formalism.

\section{The effect of a supernova explosion on protoplanetary disks}
\label{AppendixB}
\label{Sect:supernova}

A supernova may have a profound effect on proto-planetary disks in its
vicinity. Extensive simulations of this effect are performed by
\cite{2007ApJ...662.1268O,2010ApJ...711..597O}.  Similar calculations
performed by \cite{2018arXiv180204360P} included the effect of a
radiative transfer, by bridging a radiative transfer solver with a
hydrodynamics code to measure the effects of the irradiative heating
and the subsequent blast wave.

The hydrodynamics is addressed with the smoothed particles
hydrodynamics code {\tt Fi}
\citep{1989ApJS...70..419H,1997A&A...325..972G,2004A&A...422...55P},
using $10^5$ SPH particles in the disk.  The radiative transfer
calculations are performed with {\tt SPHRay}
\citep{2006PhRvE..74b6704R,2008MNRAS.386.1931A,2011ascl.soft03009A},
using $10^6$ rays per time step. A self-consistent solution is
obtained by coupling the two codes using the interlaced bridge
technique \citep{AMUSE}.

For the disks, we adopt a power-law density profile with exponent $-1$
(with temperature profile $\propto r^{-0.5}$) with Savronov-Toomre
Q-parameter $q_{\rm out} = 25$
\citep{1960AnAp...23..979S,1964ApJ...139.1217T}, for which the disk is
everywhere at least marginally stable.  The temperature of this disk
ranges from 19\,K (at the rim) to about 165\,K, in the central
regions.  We perform a series of such calculations in a grid in
distance between the supernova and the proto-planetary disk of
$d=0.05$\,pc, 0.1, 0.15, 0.2, 0.4 and 0.6\,pc and with an incident
angle with respect to the disk's normal of $\Theta = 15^\circ$, 30,
45, 60 and $75^\circ$.

\subsection{The effect of irradiation by the supernova}
\label{Sect:Irradiation}

We adopted the supernova PS1-aof11 with an energy spectrum of the
photons representative for a power-law with an index of $-3$.  The
peak of luminosity of $1.1 \times 10^{43}$\,erg/s (almost
$10^{9.8}$\,\LSun) is reached about 26 days after the supernova
explosion, producing a mass in the ejecta of 23.5\,\MSun\,
\citep{2015ApJ...799..208S}. 

\begin{figure}
\centering
\includegraphics[width=0.95\columnwidth]{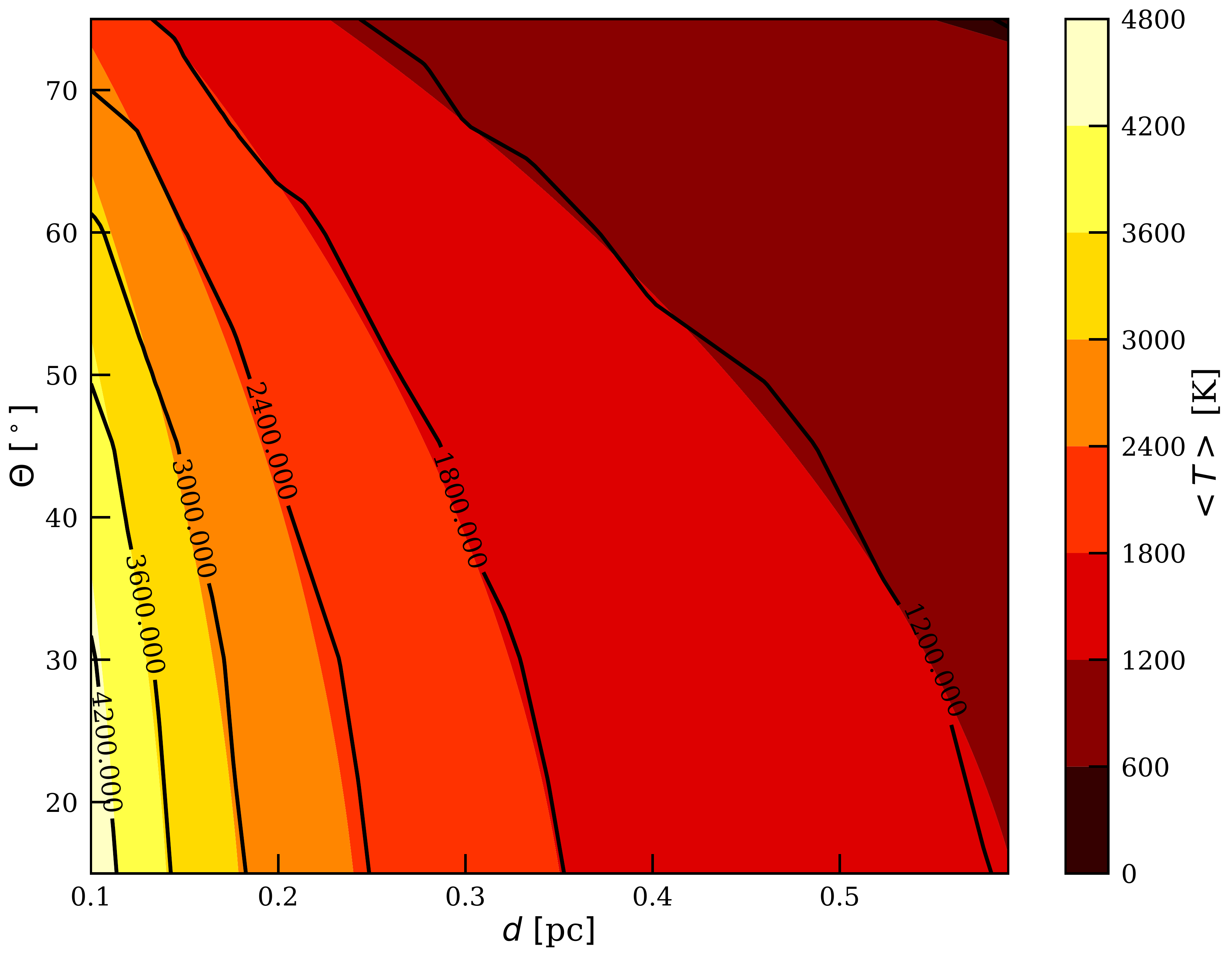}
\caption{Temperature of the disk due to the irradiation of supernova
  PS1-11aof as calculated in \cite{2018arXiv180204360P} at a distance
  $d$ and with incident angle $\Theta$. The shaded colors in the
  background give the result of Eq.\,\ref{Eq:disk_temperature}, The
  contours give the results from the grid of radiative-hydro
  calculations presented in \cite{2018arXiv180204360P}.
  \label{fig:disk_temperature}
}
\end{figure}

Protoplanetary disks in the vicinity of a supernova will be heated by
the radiation.  We fitted the result of the calculations discussed
above (see also \citep{2018arXiv180204360P}).  A satisfactory fit, at a
10\% accuracy over the entire range in incident angle $\Theta$ and
distance to the supernova $d$ was obtained
\begin{equation}
  \log_{10}(T/K) \simeq  16.47 - 13.6 d^{0.03}  \cos(\Theta)^{-0.02}.
  \label{Eq:disk_temperature}\end{equation}
The fit was carried out using the Levenberg-Marquardt algorithm
\citep{citeulike:10796881,Marq63} with the data obtained for the
calculated grid as input.  In fig.\,\ref{fig:disk_temperature} we
present the mean disk temperature from the simulations overplotted
with the fit in Eq.\,\ref{Eq:disk_temperature}

\subsection{The effect of the blast-wave impact}
\label{Sect:blastwave}

After the disk has cooled again, the nuclear blast-wave from the
supernova hits and ablates the disk, tilt it, and enriches the
surviving disk with a small amount of short-lived radionuclides. We
calculate the effect of the supernova blastwave on the circumstellar
disk by means of hydrodynamical simulations, which we perform using
the smoothed particles hydrodynamics code {\tt Fi}. Both the blastwave
as well as the disk are simulated using the same code. We adopted a
resolution of $10^5$ SPH particles in the disk as well as for the
supernova blast wave.

All hydrodynamical calculations ware performed using supernova
PS1-11aof.  For the supernova light curve, we adopt the multiple
power-law fits to the observed supernova by \cite{2015ApJ...799..208S}.
We use fit parameters for supernovae PS1-11aof from their Table\,3.
The peaks at a luminosity of $1.1 \times 10^{43}$\,erg/s (almost
$10^{9.8}$\,\LSun) about 26 days after the supernova explosion is
initiated. From the fits to the light curves,
\cite{2015ApJ...799..208S} derive masses in the ejecta of
23.5\,\MSun.

This supernova has a rather massive ejecta and we scale its effect on
the protoplanetary disk with the amount of mass in the ejecta of the
exploding star in the simulations as obtained from the stellar
evolution code.  We therefore multiply the various derived parameters
with the relative ratio of the amount of mass in the ejecta of
PS1-a11aof with respect to the amount of mass in the supernova in our
simulations $m_{\rm shell}$.  Instead of simulating the entire
impacting shell, we use only the intersecting part of the shell, with
a diameter, at the location of the disk, of twice the disk diameter.

\begin{figure}
\centering
\includegraphics[width=0.95\columnwidth]{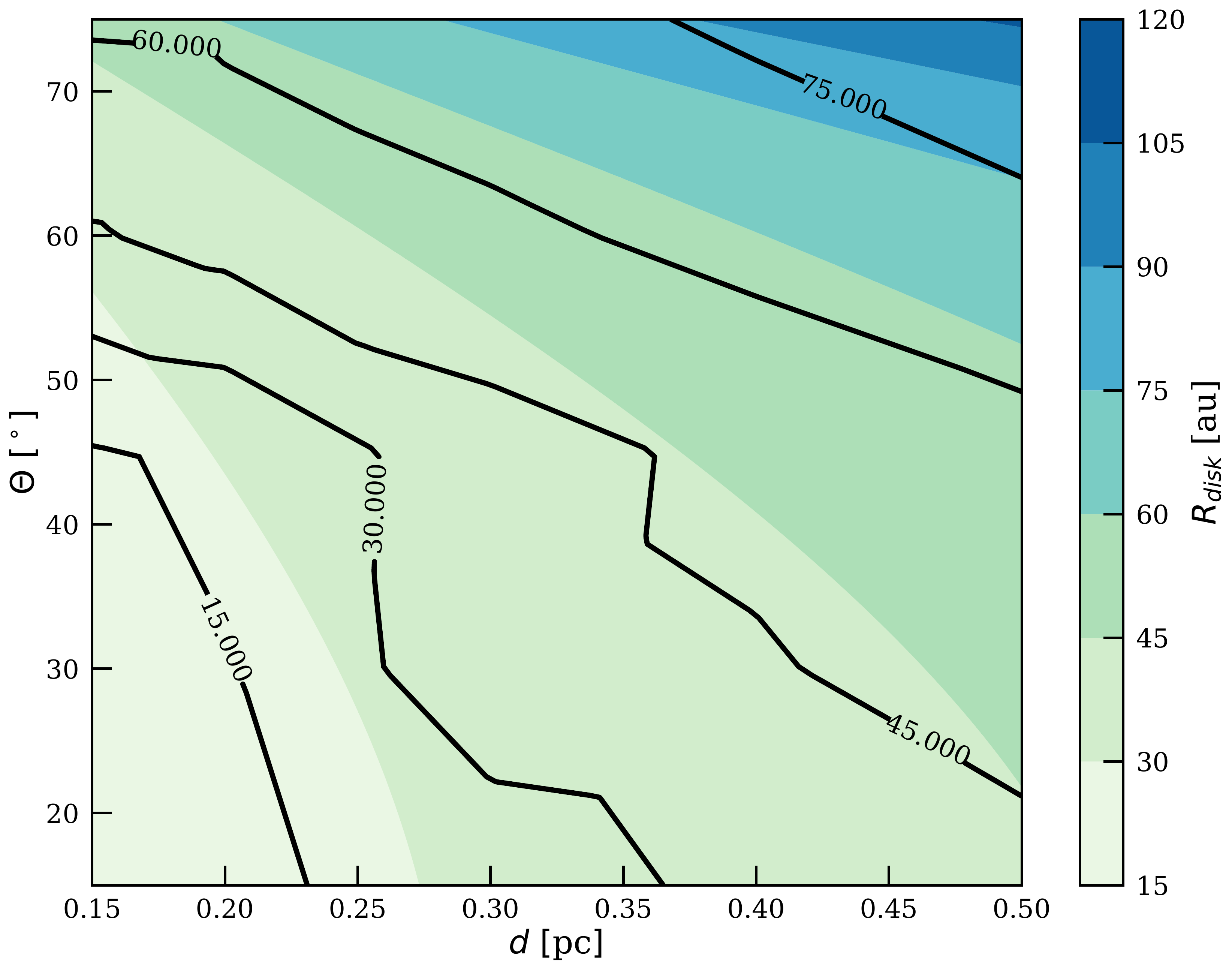}
\caption{Disk truncation due to the supernova impact (type PS1-11aof)
  as calculated in \citep{2018arXiv180204360P} in a grid in $d$ between
  0.1\,pc and 0.6\,pc and an incident angle of $\Theta = 15^\circ$ to
  $\Theta = 75^\circ$ (contours).  The shaded colors in the background
  give the result of the fit Eq.\,\ref{Eq:disk_truncation}.
  \label{fig:fit_disksize_SN11aof}
}
\end{figure}

The effect of the truncation of the disk is calculated in the same
grid of parameters as the mean temperature calculations \citep[see
  \ref{Sect:Irradiation} and also][]{2018arXiv180204360P}. We measured
disk sizes 2 years after the first contact between the blast wave and
the disk, which gives the blast wave sufficient time to pass the
entire disk.  The post-supernova disk-size was measured at a distance
from the star where the surface density of the disk drops below
2\,g/cm$^2$. We subsequently fit the disk-size as a function of the
distance $d$ and incident angle $\Theta$ with respect to the
supernova, which results in
\begin{equation}
  r_{\rm trunc}/{\rm au} \simeq 66 d^{0.63} \cos(\Theta)^{-0.68}.
\end{equation}
The fit, presented in Fig.\,\ref{fig:fit_disksize_SN11aof}, is
considerably less satisfactorily than the measured disk temperature in
Fig.\,\ref{fig:disk_temperature}. In particular with a distance to the
supernova between 0.25 and 0.45\,pc at an incident angle of $\Theta
\aplt 45^\circ$ the fit seems to break down. This is a result of the
dramatic ablation when the supernova shell hits the disk almost
face-on from a short distance. These nearby low-incident angle impacts
tend to lead to a sudden drop in disk density at the measured
distance, with a rather low surface-density disk extending to
considerable distance from the star.
The eventual disk size is calculated with
\begin{equation}
  r_{\rm disk} = r_{\rm trunc} \left( {23.5\MSun \over m_{\rm shell}} \right).
  \label{Eq:disk_truncation}
\end{equation}
Fig.\,\ref{fig:fit_disksize_SN11aof} we present the disk-size
resulting from the calculations using this supernova as a function of
incident angle and distance in Fig.\,\ref{fig:fit_disksize_SN11aof}.

\begin{figure}
\centering
\includegraphics[width=0.95\columnwidth]{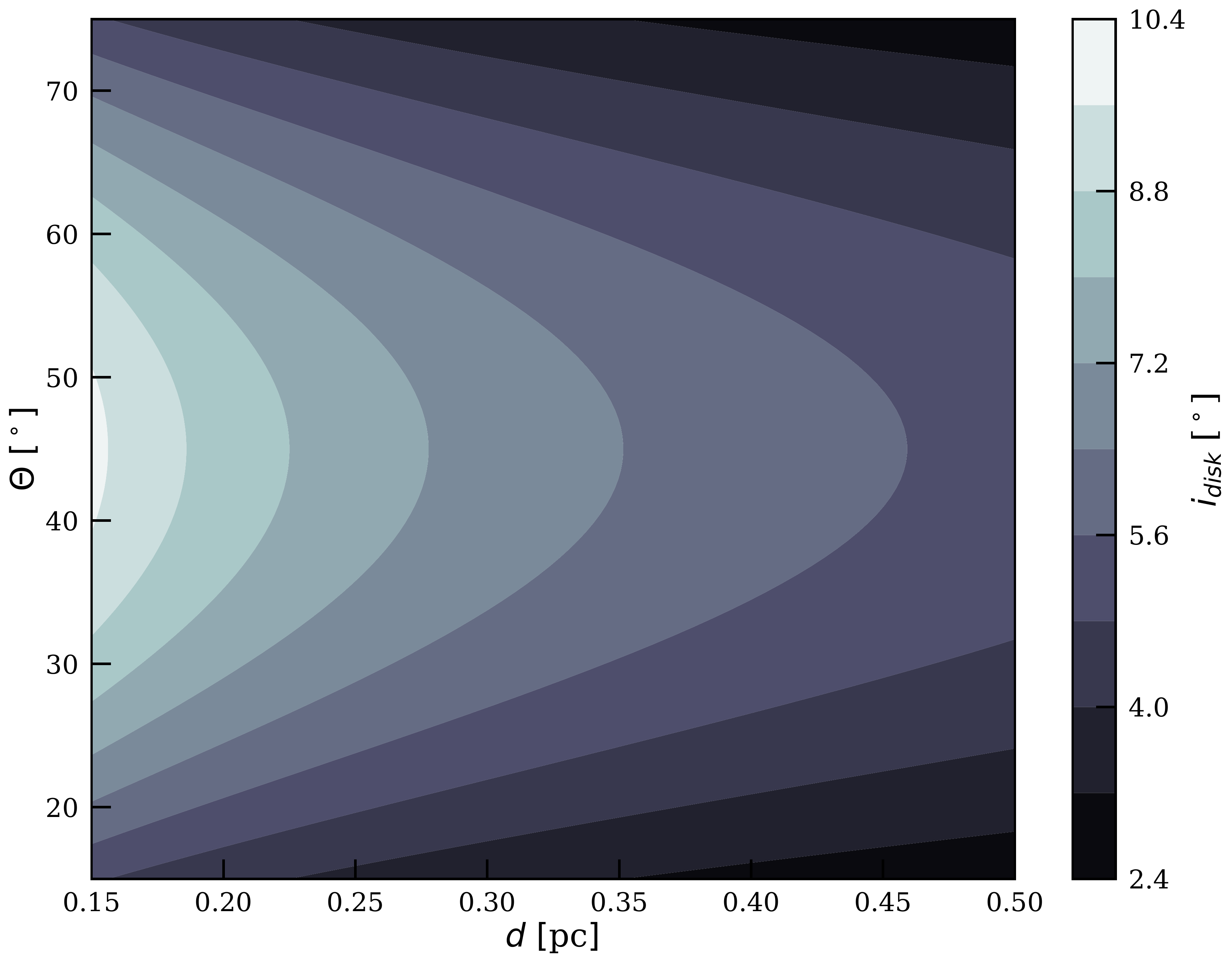}
\caption{Inclination induced upon the proto-planetary disk due to the
  supernova blast wave. The shaded colors in the background give the
  result of fit Eq.\,\ref{Eq:disk_inclination}.
  \label{fig:disk_inclination}
}
\end{figure}

As discussed in \cite{2017A&A...604A..88W} a disk hit by an external
wind will readjust itself perpendicular to the wind. This effect is
relevant for changing the inclination angle of a disk when interacting
with the supernova blast-wave.  The inclination induced by the
supernova blastwave onto the disk is calculated by the median
inclination of the truncated disk for each of the simulations in the
grid of models \citep{2018arXiv180204360P}. The fit results in
\begin{equation}
  \delta i \simeq 3.8^\circ d^{-0.50} \cos(2\Theta - 0.5\pi).
  \label{Eq:disk_inclination}
\end{equation}
In Fig.\,\ref{fig:disk_inclination} We present the resulting
inclination as a function of distance $d$ and incident angle $\Theta$
for supernova PS1-11aof.
  
\subsection{The accretion of \Fe\, and \Al\, from the supernova blast-wave}

The supernova produces \Al\, and \Fe, liberated in the supernova
blast-wave and some may be accreted by the surviving disk.  The yield
of \Fe\, in the supernova as a function of the progenitor mass is
taken from Fig.\,4\, of \citep{1995ApJ...449..204T} and can be fitted
by
\begin{equation}
  \log_{10}(m_{Fe}/\MSun) =  1.74\log_{10}(m/\MSun) - 6.93.
\end{equation}
For a 20\,\MSun\, star this results in a yield of $\sim 2.2\times
10^{-5}$\,\MSun\, in \Fe.  For \Al\, the yields are
\begin{equation}
  \log_{10}(m_{Al}/\MSun) = 2.43\log10(m/\MSun) - 7.23.
\end{equation}

\begin{figure}
\centering
\includegraphics[width=0.95\columnwidth]{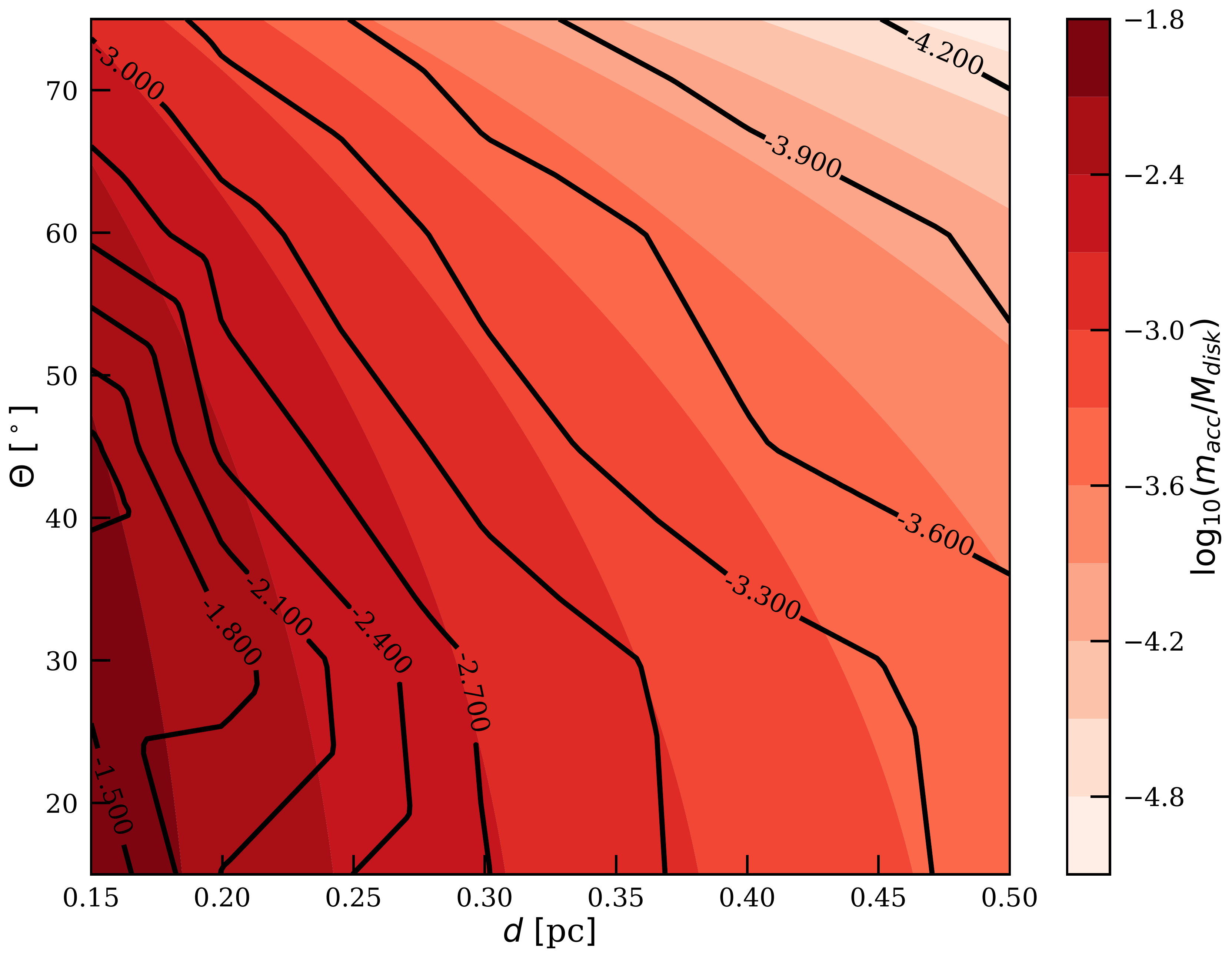}
\caption{The amount of mass that is accreted by a disk of 100\,au, as
  calculated in \cite{2018arXiv180204360P} in a grid in $d$ between
  0.1\,pc and 0.6\,pc and an incident angle $\Theta = 15^\circ$ to
  $\theta = 75^\circ$ (contours).  The shaded colors in the background
  give the result of fit Eq.\,\ref{Eq:Fe_accretion}.
  \label{fig:disk_accretion}
}
\end{figure}

The fraction of mass that is accreted by the disk from a 23.5\,\MSun\,
supernova shell was calculated by \cite{2018arXiv180204360P},
and we fitted their results to
\begin{equation}
  \log_{10}(f_{\rm acc}) = -4.77 d^{0.49} \cos(\Theta)^{-0.29}.
  \label{Eq:Fe_accretion}
\end{equation}
In fig.\ref{fig:disk_accretion} we present the fraction of mass that
is accreted by a protoplanetary disk of 100\,au\, at distance $d$ and
incident angle $\Theta$ from suprenova PS1-11aof. The actual mass of
the supernova shell that is accreted by a disk in the simulations is
calculated from the supernova shell that intersects with the disk.
Here we take the disk ($r_{\rm disk}$) into account in order to
calculate the actual amount of accreted \Fe.

The fit in Fig.\,\ref{fig:disk_accretion}, agrees within a factor of a
few with the simulations. It is hard to improve these without
performing a new extensive grid of simulations that cover a wider
range of parameters, has a higher resolution and take a wider range in
disk morphologies and supernovae into account.


\begin{thebibliography}{104}
\expandafter\ifx\csname natexlab\endcsname\relax\def\natexlab#1{#1}\fi

\bibitem[{{Adams}(2010)}]{2010ARA&A..48...47A}
{Adams}, F.~C. 2010, \araa, 48, 47

\bibitem[{{Allen}(1973)}]{1973asqu.book.....A}
{Allen}, C.~W. 1973, {Astrophysical quantities}

\bibitem[{{Altay} {et~al.}(2008){Altay}, {Croft}, \&
  {Pelupessy}}]{2008MNRAS.386.1931A}
{Altay}, G., {Croft}, R.~A.~C., \& {Pelupessy}, I. 2008, \mnras, 386, 1931

\bibitem[{{Altay} {et~al.}(2011){Altay}, {Croft}, \&
  {Pelupessy}}]{2011ascl.soft03009A}
{Altay}, G., {Croft}, R.~A.~C., \& {Pelupessy}, I. 2011, {SPHRAY: A Smoothed
  Particle Hydrodynamics Ray Tracer for Radiative Transfer}, Astrophysics
  Source Code Library

\bibitem[{Arakawa \& Nakamoto(2016)}]{ARAKAWA2016102}
Arakawa, S. \& Nakamoto, T. 2016, Icarus, 276, 102

\bibitem[{{Barenfeld} {et~al.}(2017){Barenfeld}, {Carpenter}, {Sargent},
  {Isella}, \& {Ricci}}]{2017ApJ...851...85B}
{Barenfeld}, S.~A., {Carpenter}, J.~M., {Sargent}, A.~I., {Isella}, A., \&
  {Ricci}, L. 2017, \apj, 851, 85

\bibitem[{{Beer} {et~al.}(2004){Beer}, {King}, {Livio}, \&
  {Pringle}}]{2004MNRAS.354..763B}
{Beer}, M.~E., {King}, A.~R., {Livio}, M., \& {Pringle}, J.~E. 2004, \mnras,
  354, 763

\bibitem[{Bondi \& Hoyle(1944)}]{bondy_hoyle44}
Bondi, H. \& Hoyle, F. 1944, \mnras, 104, 273

\bibitem[{{Breslau} {et~al.}(2014){Breslau}, {Steinhausen}, {Vincke}, \&
  {Pfalzner}}]{2014A&A...565A.130B}
{Breslau}, A., {Steinhausen}, M., {Vincke}, K., \& {Pfalzner}, S. 2014, \aap,
  565, A130

\bibitem[{{Cai} {et~al.}(2018){Cai}, {Portegies Zwart}, \& {van
  Elteren}}]{2018MNRAS.474.5114C}
{Cai}, M.~X., {Portegies Zwart}, S., \& {van Elteren}, A. 2018, \mnras, 474,
  5114

\bibitem[{{Davis} \& {Richter}(2005)}]{2005mcp..book..407D}
{Davis}, A.~M. \& {Richter}, F.~M. 2005, {Condensation and Evaporation of Solar
  System Materials}, ed. A.~M. {Davis}, H.~D. {Holland}, \& K.~K. {Turekian}
  (Elsevier B), 407

\bibitem[{{Dearborn} \& {Blake}(1988)}]{1988ApJ...332..305D}
{Dearborn}, D.~S.~P. \& {Blake}, J.~B. 1988, \apj, 332, 305

\bibitem[{{Drimmel}(2000)}]{2000A&A...358L..13D}
{Drimmel}, R. 2000, \aap, 358, L13

\bibitem[{{Dwarkadas} {et~al.}(2017){Dwarkadas}, {Dauphas}, {Meyer},
  {Boyajian}, \& {Bojazi}}]{2017ApJ...851..147D}
{Dwarkadas}, V.~V., {Dauphas}, N., {Meyer}, B., {Boyajian}, P., \& {Bojazi}, M.
  2017, \apj, 851, 147

\bibitem[{{Eisner} {et~al.}(2018){Eisner}, {Arce}, {Ballering}, {Bally},
  {Andrews}, {Boyden}, {Di Francesco}, {Fang}, {Johnstone}, {Kim}, {Mann},
  {Matthews}, {Pascucci}, {Ricci}, {Sheehan}, \&
  {Williams}}]{2018ApJ...860...77E}
{Eisner}, J.~A., {Arce}, H.~G., {Ballering}, N.~P., {et~al.} 2018, \apj, 860,
  77

\bibitem[{{Fujii} {et~al.}(2007){Fujii}, {Iwasawa}, {Funato}, \&
  {Makino}}]{2007PASJ...59.1095F}
{Fujii}, M., {Iwasawa}, M., {Funato}, Y., \& {Makino}, J. 2007, \pasj, 59, 1095

\bibitem[{{Gaidos} {et~al.}(2009){Gaidos}, {Krot}, {Williams}, \&
  {Raymond}}]{2009ApJ...696.1854G}
{Gaidos}, E., {Krot}, A.~N., {Williams}, J.~P., \& {Raymond}, S.~N. 2009, \apj,
  696, 1854

\bibitem[{{Galilei}(1632)}]{1632Dialogo...G}
{Galilei}, G. 1632 ({Dello Studio di Pisa}), 189--225

\bibitem[{{Gerhard}(2011)}]{2011MSAIS..18..185G}
{Gerhard}, O. 2011, Memorie della Societa Astronomica Italiana Supplementi, 18,
  185

\bibitem[{{Gerritsen} \& {Icke}(1997)}]{1997A&A...325..972G}
{Gerritsen}, J.~P.~E. \& {Icke}, V. 1997, \aap, 325, 972

\bibitem[{{Goodwin} \& {Whitworth}(2004)}]{2004A&A...413..929G}
{Goodwin}, S.~P. \& {Whitworth}, A.~P. 2004, \aap, 413, 929

\bibitem[{{Gounelle}(2015)}]{2015AA...582A..26G}
{Gounelle}, M. 2015, \aap, 582, A26

\bibitem[{{Gounelle} \& {Meynet}(2012)}]{gounelle12}
{Gounelle}, M. \& {Meynet}, G. 2012, \aap, 545, A4

\bibitem[{{Hernquist} \& {Katz}(1989)}]{1989ApJS...70..419H}
{Hernquist}, L. \& {Katz}, N. 1989, \apjs, 70, 419

\bibitem[{{Hewins} \& {Radomsky}(1990)}]{1990Metic..25..309H}
{Hewins}, R.~H. \& {Radomsky}, P.~M. 1990, Meteoritics, 25, 309

\bibitem[{Horányi {et~al.}(1995)Horányi, Morfill, Goertz, \&
  Levy}]{HORANYI1995174}
Horányi, M., Morfill, G., Goertz, C., \& Levy, E. 1995, Icarus, 114, 174

\bibitem[{{Hsu} {et~al.}(2013){Hsu}, {Hartmann}, {Allen}, {Hern{\'a}ndez},
  {Megeath}, {Tobin}, \& {Ingleby}}]{2013ApJ...764..114H}
{Hsu}, W.-H., {Hartmann}, L., {Allen}, L., {et~al.} 2013, \apj, 764, 114

\bibitem[{{Isella} {et~al.}(2009){Isella}, {Carpenter}, \&
  {Sargent}}]{2009ApJ...701..260I}
{Isella}, A., {Carpenter}, J.~M., \& {Sargent}, A.~I. 2009, \apj, 701, 260

\bibitem[{{Jacobsen} {et~al.}(2008){Jacobsen}, {Chakrabarti}, {Ranen}, \&
  {Petaev}}]{2008LPI....39.1999J}
{Jacobsen}, S.~B., {Chakrabarti}, R., {Ranen}, M.~C., \& {Petaev}, M.~I. 2008,
  in Lunar and Planetary Science Conference, Vol.~39, Lunar and Planetary
  Science Conference, 1999

\bibitem[{{J{\'{\i}}lkov{\'a}} {et~al.}(2012){J{\'{\i}}lkov{\'a}}, {Carraro},
  {Jungwiert}, \& {Minchev}}]{2012A&A...541A..64J}
{J{\'{\i}}lkov{\'a}}, L., {Carraro}, G., {Jungwiert}, B., \& {Minchev}, I.
  2012, \aap, 541, A64

\bibitem[{{J{\'{\i}}lkov{\'a}} {et~al.}(2016){J{\'{\i}}lkov{\'a}}, {Hamers},
  {Hammer}, \& {Portegies Zwart}}]{2016MNRAS.457.4218J}
{J{\'{\i}}lkov{\'a}}, L., {Hamers}, A.~S., {Hammer}, M., \& {Portegies Zwart},
  S. 2016, \mnras, 457, 4218

\bibitem[{{Juri{\'c}} {et~al.}(2008){Juri{\'c}}, {Ivezi{\'c}}, {Brooks},
  {Lupton}, {Schlegel}, {Finkbeiner}, {Padmanabhan}, {Bond}, {Sesar},
  {Rockosi}, {Knapp}, {Gunn}, {Sumi}, {Schneider}, {Barentine}, {Brewington},
  {Brinkmann}, {Fukugita}, {Harvanek}, {Kleinman}, {Krzesinski}, {Long},
  {Neilsen}, {Nitta}, {Snedden}, \& {York}}]{2008ApJ...673..864J}
{Juri{\'c}}, M., {Ivezi{\'c}}, {\v Z}., {Brooks}, A., {et~al.} 2008, \apj, 673,
  864

\bibitem[{{Kapteyn}(1922)}]{1922ApJ....55..302K}
{Kapteyn}, J.~C. 1922, \apj, 55, 302

\bibitem[{{Kouwenhoven} {et~al.}(2014){Kouwenhoven}, {Goodwin}, {de Grijs},
  {Rose}, \& {Kim}}]{2014MNRAS.445.2256K}
{Kouwenhoven}, M.~B.~N., {Goodwin}, S.~P., {de Grijs}, R., {Rose}, M., \&
  {Kim}, S.~S. 2014, \mnras, 445, 2256

\bibitem[{{Kroupa}(2002)}]{2002Sci...295...82K}
{Kroupa}, P. 2002, Science, 295, 82

\bibitem[{{Kroupa} \& {Weidner}(2003)}]{2003ApJ...598.1076K}
{Kroupa}, P. \& {Weidner}, C. 2003, \apj, 598, 1076

\bibitem[{{Krumholz}(2014)}]{2014PhR...539...49K}
{Krumholz}, M.~R. 2014, \physrep, 539, 49

\bibitem[{{Krumholz} {et~al.}(2015){Krumholz}, {Fumagalli}, {da Silva},
  {Rendahl}, \& {Parra}}]{2015MNRAS.452.1447K}
{Krumholz}, M.~R., {Fumagalli}, M., {da Silva}, R.~L., {Rendahl}, T., \&
  {Parra}, J. 2015, \mnras, 452, 1447

\bibitem[{Levenberg(1944)}]{citeulike:10796881}
Levenberg, K. 1944, Quarterly Journal of Applied Mathmatics, II, 164

\bibitem[{{Lichtenberg} {et~al.}(2016){Lichtenberg}, {Parker}, \&
  {Meyer}}]{2016MNRAS.462.3979L}
{Lichtenberg}, T., {Parker}, R.~J., \& {Meyer}, M.~R. 2016, \mnras, 462, 3979

\bibitem[{{Lodders}(2003)}]{2003ApJLodders}
{Lodders}, K. 2003, \apj, 591, 1220

\bibitem[{{Lyo} \& {Lawson}(2005)}]{2005JKAS...38..241L}
{Lyo}, A.-R. \& {Lawson}, W.~A. 2005, Journal of Korean Astronomical Society,
  38, 241

\bibitem[{{Lyo} {et~al.}(2003){Lyo}, {Lawson}, {Mamajek}, {Feigelson}, {Sung},
  \& {Crause}}]{2003MNRAS.338..616L}
{Lyo}, A.-R., {Lawson}, W.~A., {Mamajek}, E.~E., {et~al.} 2003, \mnras, 338,
  616

\bibitem[{{MacPherson} {et~al.}(1995){MacPherson}, {Davis}, \&
  {Zinner}}]{1995Metic..30..365M}
{MacPherson}, G.~J., {Davis}, A.~M., \& {Zinner}, E.~K. 1995, Meteoritics, 30,
  365

\bibitem[{{Mamajek} {et~al.}(2004){Mamajek}, {Meyer}, {Hinz}, {Hoffmann},
  {Cohen}, \& {Hora}}]{2004ApJ...612..496M}
{Mamajek}, E.~E., {Meyer}, M.~R., {Hinz}, P.~M., {et~al.} 2004, \apj, 612, 496

\bibitem[{Marquardt(1963)}]{Marq63}
Marquardt, D.~W. 1963, Journal of the Society for Industrial and Applied
  Mathematics, 11, 431

\bibitem[{{Mart{\'{\i}}nez-Barbosa} {et~al.}(2016){Mart{\'{\i}}nez-Barbosa},
  {Brown}, {Boekholt}, {Portegies Zwart}, {Antiche}, \&
  {Antoja}}]{2016MNRAS.457.1062M}
{Mart{\'{\i}}nez-Barbosa}, C.~A., {Brown}, A.~G.~A., {Boekholt}, T., {et~al.}
  2016, \mnras, 457, 1062

\bibitem[{{Mart{\'{\i}}nez-Barbosa} {et~al.}(2015){Mart{\'{\i}}nez-Barbosa},
  {Brown}, \& {Portegies Zwart}}]{2015MNRAS.446..823M}
{Mart{\'{\i}}nez-Barbosa}, C.~A., {Brown}, A.~G.~A., \& {Portegies Zwart}, S.
  2015, \mnras, 446, 823

\bibitem[{{Mart{\'{\i}}nez-Barbosa} {et~al.}(2017){Mart{\'{\i}}nez-Barbosa},
  {J{\'{\i}}lkov{\'a}}, {Portegies Zwart}, \& {Brown}}]{2017MNRAS.464.2290M}
{Mart{\'{\i}}nez-Barbosa}, C.~A., {J{\'{\i}}lkov{\'a}}, L., {Portegies Zwart},
  S., \& {Brown}, A.~G.~A. 2017, \mnras, 464, 2290

\bibitem[{{Mishra} {et~al.}(2016){Mishra}, {Marhas}, \&
  {Sameer}}]{2016E&PSL.436...71M}
{Mishra}, R.~K., {Marhas}, K.~K., \& {Sameer}. 2016, Earth and Planetary
  Science Letters, 436, 71

\bibitem[{{Monari} {et~al.}(2014){Monari}, {Helmi}, {Antoja}, \&
  {Steinmetz}}]{2014A&A...569A..69M}
{Monari}, G., {Helmi}, A., {Antoja}, T., \& {Steinmetz}, M. 2014, \aap, 569,
  A69

\bibitem[{{Mowlavi} \& {Meynet}(2000)}]{2000A&A...361..959M}
{Mowlavi}, N. \& {Meynet}, G. 2000, \aap, 361, 959

\bibitem[{{Nicholson} \& {Parker}(2017)}]{2017MNRAS.464.4318N}
{Nicholson}, R.~B. \& {Parker}, R.~J. 2017, \mnras, 464, 4318

\bibitem[{{Nomoto} {et~al.}(2006){Nomoto}, {Tominaga}, {Umeda}, {Kobayashi}, \&
  {Maeda}}]{2006NuPhA.777..424N}
{Nomoto}, K., {Tominaga}, N., {Umeda}, H., {Kobayashi}, C., \& {Maeda}, K.
  2006, Nuclear Physics A, 777, 424

\bibitem[{{Ouellette} {et~al.}(2007){Ouellette}, {Desch}, \&
  {Hester}}]{2007ApJ...662.1268O}
{Ouellette}, N., {Desch}, S.~J., \& {Hester}, J.~J. 2007, \apj, 662, 1268

\bibitem[{{Ouellette} {et~al.}(2010){Ouellette}, {Desch}, \&
  {Hester}}]{2010ApJ...711..597O}
{Ouellette}, N., {Desch}, S.~J., \& {Hester}, J.~J. 2010, \apj, 711, 597

\bibitem[{{Parker} {et~al.}(2014){Parker}, {Church}, {Davies}, \&
  {Meyer}}]{2014MNRAS.437..946P}
{Parker}, R.~J., {Church}, R.~P., {Davies}, M.~B., \& {Meyer}, M.~R. 2014,
  \mnras, 437, 946

\bibitem[{{Parker} \& {Dale}(2016)}]{2016MNRAS.456.1066P}
{Parker}, R.~J. \& {Dale}, J.~E. 2016, \mnras, 456, 1066

\bibitem[{{Parker} \& {Quanz}(2012)}]{2012MNRAS.419.2448P}
{Parker}, R.~J. \& {Quanz}, S.~P. 2012, \mnras, 419, 2448

\bibitem[{{Pelupessy} {et~al.}(2004){Pelupessy}, {van der Werf}, \&
  {Icke}}]{2004A&A...422...55P}
{Pelupessy}, F.~I., {van der Werf}, P.~P., \& {Icke}, V. 2004, \aap, 422, 55

\bibitem[{{Pelupessy} {et~al.}(2013){Pelupessy}, {van Elteren}, {de Vries},
  {McMillan}, {Drost}, \& {Portegies Zwart}}]{2013AA...557A..84P}
{Pelupessy}, F.~I., {van Elteren}, A., {de Vries}, N., {et~al.} 2013, \aap,
  557, A84

\bibitem[{{Pfalzner} {et~al.}(2014){Pfalzner}, {Steinhausen}, \&
  {Menten}}]{2014ApJ...793L..34P}
{Pfalzner}, S., {Steinhausen}, M., \& {Menten}, K. 2014, \apjl, 793, L34

\bibitem[{{Plummer}(1911)}]{1911MNRAS..71..460P}
{Plummer}, H.~C. 1911, \mnras, 71, 460

\bibitem[{{Portegies Zwart}(2011)}]{2011ascl.soft07007P}
{Portegies Zwart}, S. 2011, {AMUSE: Astrophysical Multipurpose Software
  Environment}, Astrophysics Source Code Library

\bibitem[{{Portegies Zwart} \& Boekholt(2014)}]{2041-8205-785-1-L3}
{Portegies Zwart}, S. \& Boekholt, T. 2014, The Astrophysical Journal Letters,
  785, L3

\bibitem[{{Portegies Zwart} \& {McMillan}(2018)}]{AMUSE}
{Portegies Zwart}, S. \& {McMillan}, S. 2018, {Astrophysical Recipes: the Art
  of AMUSE} (AAS IOP Astronomy ({\em in press}))

\bibitem[{{Portegies Zwart} {et~al.}(2018){Portegies Zwart}, {Pelupessy}, {van
  Elteren}, {Wijnen}, \& {Lugaro}}]{2018arXiv180204360P}
{Portegies Zwart}, S., {Pelupessy}, I., {van Elteren}, A., {Wijnen}, T., \&
  {Lugaro}, M. 2018, ArXiv e-prints [\eprint[arXiv]{1802.04360}]

\bibitem[{{Portegies Zwart}(2009)}]{2009ApJ...696L..13P}
{Portegies Zwart}, S.~F. 2009, \apjl, 696, L13

\bibitem[{{Portegies Zwart}(2016)}]{2016MNRAS.457..313P}
{Portegies Zwart}, S.~F. 2016, \mnras, 457, 313

\bibitem[{{Portegies Zwart} {et~al.}(2013){Portegies Zwart}, McMillan, van
  Elteren, Pelupessy, \& de~Vries}]{PortegiesZwart2013456}
{Portegies Zwart}, S.~F., McMillan, S.~L., van Elteren, A., Pelupessy, F.~I.,
  \& de~Vries, N. 2013, Computer Physics Communications, 184, 456

\bibitem[{{Portegies Zwart} {et~al.}(2010){Portegies Zwart}, {McMillan}, \&
  {Gieles}}]{2010ARA&A..48..431P}
{Portegies Zwart}, S.~F., {McMillan}, S.~L.~W., \& {Gieles}, M. 2010, \araa,
  48, 431

\bibitem[{{Portegies Zwart} {et~al.}(2001){Portegies Zwart}, {McMillan}, {Hut},
  \& {Makino}}]{2001MNRAS.321..199P}
{Portegies Zwart}, S.~F., {McMillan}, S.~L.~W., {Hut}, P., \& {Makino}, J.
  2001, \mnras, 321, 199

\bibitem[{{Portegies Zwart} \& {Verbunt}(1996)}]{1996A&A...309..179P}
{Portegies Zwart}, S.~F. \& {Verbunt}, F. 1996, \aap, 309, 179

\bibitem[{{Portegies Zwart} \& {Verbunt}(2012)}]{2012ascl.soft01003P}
{Portegies Zwart}, S.~F. \& {Verbunt}, F. 2012, {SeBa: Stellar and binary
  evolution}, Astrophysics Source Code Library

\bibitem[{{Portegies Zwart} \& {Yungelson}(1998)}]{1998A&A...332..173P}
{Portegies Zwart}, S.~F. \& {Yungelson}, L.~R. 1998, \aap, 332, 173

\bibitem[{{Punzo} {et~al.}(2014){Punzo}, {Capuzzo-Dolcetta}, \& {Portegies
  Zwart}}]{2014MNRAS.444.2808P}
{Punzo}, D., {Capuzzo-Dolcetta}, R., \& {Portegies Zwart}, S. 2014, \mnras,
  444, 2808

\bibitem[{{Reddish}(1978)}]{1978stfo.book.....R}
{Reddish}, V.~C. 1978, {Stellar formation}

\bibitem[{{Ribas} {et~al.}(2015){Ribas}, {Bouy}, \&
  {Mer{\'{\i}}n}}]{2015A&A...576A..52R}
{Ribas}, {\'A}., {Bouy}, H., \& {Mer{\'{\i}}n}, B. 2015, \aap, 576, A52

\bibitem[{{Richert} {et~al.}(2018){Richert}, {Getman}, {Feigelson}, {Kuhn},
  {Broos}, {Povich}, {Bate}, \& {Garmire}}]{2018MNRAS.477.5191R}
{Richert}, A.~J.~W., {Getman}, K.~V., {Feigelson}, E.~D., {et~al.} 2018,
  \mnras, 477, 5191

\bibitem[{{Ritzerveld} \& {Icke}(2006)}]{2006PhRvE..74b6704R}
{Ritzerveld}, J. \& {Icke}, V. 2006, Phys.\, Rev.\, E, 74, 026704

\bibitem[{{Robitaille} \& {Whitney}(2010)}]{2010ApJ...710L..11R}
{Robitaille}, T.~P. \& {Whitney}, B.~A. 2010, \apjl, 710, L11

\bibitem[{{Romero-G{\'o}mez} {et~al.}(2011){Romero-G{\'o}mez}, {Athanassoula},
  {Antoja}, \& {Figueras}}]{2011MNRAS.418.1176R}
{Romero-G{\'o}mez}, M., {Athanassoula}, E., {Antoja}, T., \& {Figueras}, F.
  2011, \mnras, 418, 1176

\bibitem[{{Ronco} \& {de El{\'{\i}}a}(2018)}]{2018MNRAS.tmp.1696R}
{Ronco}, M.~P. \& {de El{\'{\i}}a}, G.~C. 2018, \mnras
  [\eprint[arXiv]{1807.01429}]

\bibitem[{{Ronco} {et~al.}(2017){Ronco}, {Guilera}, \& {de
  El{\'{\i}}a}}]{2017MNRAS.471.2753R}
{Ronco}, M.~P., {Guilera}, O.~M., \& {de El{\'{\i}}a}, G.~C. 2017, \mnras, 471,
  2753

\bibitem[{Rugel {et~al.}(2009)Rugel, Faestermann, Knie, Korschinek, Poutivtsev,
  Schumann, Kivel, G\"unther-Leopold, Weinreich, \&
  Wohlmuther}]{PhysRevLett.103.072502}
Rugel, G., Faestermann, T., Knie, K., {et~al.} 2009, Phys. Rev. Lett., 103,
  072502

\bibitem[{{Safronov}(1960)}]{1960AnAp...23..979S}
{Safronov}, V.~S. 1960, Annales d'Astrophysique, 23, 979

\bibitem[{Sanders \& Scott(2012)}]{SANDERS_2012}
Sanders, I.~S. \& Scott, E. R.~D. 2012, Meteoritics {\&} Planetary Science, 47,
  2170

\bibitem[{{Sanders} {et~al.}(2015){Sanders}, {Soderberg}, {Gezari},
  {Betancourt}, {Chornock}, {Berger}, {Foley}, {Challis}, {Drout}, {Kirshner},
  {Lunnan}, {Marion}, {Margutti}, {McKinnon}, {Milisavljevic}, {Narayan},
  {Rest}, {Kankare}, {Mattila}, {Smartt}, {Huber}, {Burgett}, {Draper},
  {Hodapp}, {Kaiser}, {Kudritzki}, {Magnier}, {Metcalfe}, {Morgan}, {Price},
  {Tonry}, {Wainscoat}, \& {Waters}}]{2015ApJ...799..208S}
{Sanders}, N.~E., {Soderberg}, A.~M., {Gezari}, S., {et~al.} 2015, \apj, 799,
  208

\bibitem[{{Schechter}(1976)}]{1976ApJ...203..297S}
{Schechter}, P. 1976, \apj, 203, 297

\bibitem[{{Tatischeff} {et~al.}(2010){Tatischeff}, {Duprat}, \& {de
  S{\'e}r{\'e}ville}}]{2010ApJ...714L..26T}
{Tatischeff}, V., {Duprat}, J., \& {de S{\'e}r{\'e}ville}, N. 2010, \apjl, 714,
  L26

\bibitem[{{Tazzari} {et~al.}(2017){Tazzari}, {Testi}, {Natta}, {Ansdell},
  {Carpenter}, {Guidi}, {Hogerheijde}, {Manara}, {Miotello}, {van der Marel},
  {van Dishoeck}, \& {Williams}}]{2017A&A...606A..88T}
{Tazzari}, M., {Testi}, L., {Natta}, A., {et~al.} 2017, \aap, 606, A88

\bibitem[{{Timmes} {et~al.}(1995){Timmes}, {Woosley}, {Hartmann}, {Hoffman},
  {Weaver}, \& {Matteucci}}]{1995ApJ...449..204T}
{Timmes}, F.~X., {Woosley}, S.~E., {Hartmann}, D.~H., {et~al.} 1995, \apj, 449,
  204

\bibitem[{{Toomre}(1964)}]{1964ApJ...139.1217T}
{Toomre}, A. 1964, \apj, 139, 1217

\bibitem[{{Toonen} {et~al.}(2012){Toonen}, {Nelemans}, \& {Portegies
  Zwart}}]{2012A&A...546A..70T}
{Toonen}, S., {Nelemans}, G., \& {Portegies Zwart}, S. 2012, \aap, 546, A70

\bibitem[{{Trappitsch} {et~al.}(2018){Trappitsch}, {Boehnke}, {Stephan},
  {Telus}, {Savina}, {Pardo}, {Davis}, {Dauphas}, {Pellin}, \&
  {Huss}}]{2018ApJ...857L..15T}
{Trappitsch}, R., {Boehnke}, P., {Stephan}, T., {et~al.} 2018, \apjl, 857, L15

\bibitem[{{Tripathi} {et~al.}(2017){Tripathi}, {Andrews}, {Birnstiel}, \&
  {Wilner}}]{2017ApJ...845...44T}
{Tripathi}, A., {Andrews}, S.~M., {Birnstiel}, T., \& {Wilner}, D.~J. 2017,
  \apj, 845, 44

\bibitem[{{van den Heuvel} \& {Portegies Zwart}(2013)}]{2013ApJ...779..114V}
{van den Heuvel}, E.~P.~J. \& {Portegies Zwart}, S.~F. 2013, \apj, 779, 114

\bibitem[{{Vanbeveren}(1982)}]{1982A&A...115...65V}
{Vanbeveren}, D. 1982, \aap, 115, 65

\bibitem[{{Vincke} {et~al.}(2015){Vincke}, {Breslau}, \&
  {Pfalzner}}]{2015A&A...577A.115V}
{Vincke}, K., {Breslau}, A., \& {Pfalzner}, S. 2015, \aap, 577, A115

\bibitem[{{Vincke} \& {Pfalzner}(2016)}]{2016ApJ...828...48V}
{Vincke}, K. \& {Pfalzner}, S. 2016, \apj, 828, 48

\bibitem[{{Vuissoz} {et~al.}(2004){Vuissoz}, {Meynet}, {Kn{\"o}dlseder},
  {Cervi{\~n}o}, {Schaerer}, {Palacios}, \& {Mowlavi}}]{2004NewAR..48....7V}
{Vuissoz}, C., {Meynet}, G., {Kn{\"o}dlseder}, J., {et~al.} 2004, \nar, 48, 7

\bibitem[{{Wasserburg} {et~al.}(2006){Wasserburg}, {Busso}, {Gallino}, \&
  {Nollett}}]{2006NuPhA.777....5W}
{Wasserburg}, G.~J., {Busso}, M., {Gallino}, R., \& {Nollett}, K.~M. 2006,
  Nuclear Physics A, 777, 5

\bibitem[{{Weidner} \& {Kroupa}(2006)}]{2006MNRAS.365.1333W}
{Weidner}, C. \& {Kroupa}, P. 2006, \mnras, 365, 1333

\bibitem[{{Wijnen} {et~al.}(2017){Wijnen}, {Pelupessy}, {Pols}, \& {Portegies
  Zwart}}]{2017A&A...604A..88W}
{Wijnen}, T.~P.~G., {Pelupessy}, F.~I., {Pols}, O.~R., \& {Portegies Zwart}, S.
  2017, \aap, 604, A88

\end{thebibliography}
\end{document}